\documentclass[useAMS,usenatbib]{mn2e}

\usepackage{amssymb}
\usepackage{aas_macros}
\usepackage{natbib}
\usepackage{times}
\usepackage{graphicx}

\newcommand{\figwidth}{\columnwidth }
\newcommand{\sigmag}{\Sigma_{\rm g}}
\newcommand{\Zej}{Z_{\rm ej}}
\newcommand{\Zg}{Z_{\rm g}}
\newcommand{\Mgas}{M_{\rm g}}
\newcommand{\Msun}{{\rm M}_\odot}
\newcommand{\fR}{f_{\rm R}}
\newcommand{\fg}{f_{\rm g}}
\newcommand{\FR}{F_{\rm R}}
\newcommand{\betaZ}{\beta_{\rm Z}}
\newcommand{\CTB}{\nocite{Creasey_13}Paper~I}

\newcommand{\flash}{{\sc flash}}

\newcommand{\amr}{{\sc amr}}

\title[The metallicity of galactic winds]{The metallicity of galactic winds}
\author[Creasey, et al.]  {\parbox[h]{160mm} { 
    Peter Creasey$^{1,2}$\thanks{E-mail: pcreasey@aip.de},
    Tom Theuns$^{2,3}$ and
Richard G. Bower$^2$}
  \vspace{6pt}\\
  $^1$Leibniz-Institut f{\"u}r Astrophysik Potsdam (AIP)
An der Sternwarte 16, 14482 Potsdam, Germany\\
  $^2$Institute for Computational Cosmology, Department of Physics,
  University of Durham, South Road, Durham, DH1 3LE, UK\\
  $^3$Department of Physics, University of Antwerp, Campus Groenenborger, Groenenborgerlaan 171, B-2020 Antwerp,  Belgium}

\begin{document}

\date{\today}
\pagerange{\pageref{firstpage}--\pageref{lastpage}} \pubyear{2010}

\maketitle

\label{firstpage}

\begin{abstract}
 
The abundance evolution of galaxies depends critically on the balance between the mixing of metals in their interstellar medium, the inflow of new gas and the outflow of enriched gas. We study these processes
in gas columns perpendicular to a galactic disk using sub-parsec  resolution  simulations that track stellar ejecta with the \flash\ code. We model a simplified interstellar medium stirred and enriched by supernovae and their progenitors.
We vary the density distribution of the gas column and integrate our results over an exponential disk to predict wind and ISM enrichment properties for disk galaxies. 
We find that winds from more massive galaxies are hotter and more highly enriched, in stark contrast to that which is often assumed in galaxy formation models. 
We use these findings in a simple model of galactic enrichment evolution, in which the metallicity of forming galaxies is the result of accretion of nearly pristine gas and outflow of enriched gas along an equilibrium sequence. 
We compare these predictions to the observed mass-metallicity relation, and demonstrate how the galaxy's gas fraction 
is a key controlling parameter. This explains the observed flattening of the mass-metallicity relation at higher stellar masses.

\end{abstract}

\section{Introduction}
Supernovae are a key ingredient in current models of galaxy formation. In these models, the energy associated with core-collapse \lq Type~II\rq\ supernovae (SNe) is invoked to drive a galactic wind. 
In small galaxies, such winds eject a large fraction of the baryons \citep{Larson_1974,Rees_1977, WhiteRees_1978, Benson_2003} thereby regulating their star formation \citep{Schaye_10}. Such drastic \lq feedback\rq\ is required by the models to explain the inefficiency with which gas is converted to stars as inferred from the baryon census of the universe \citep{Balogh_2001, Tumlinson_2013}, the stellar mass of galaxies of a given halo mass \citep{Guo_10, Behroozi_10}, and the presence of elements synthesised in stars (\lq metals\rq) in the intergalactic medium \citep{Theuns_02, Springel_03,Aguirre_05, Cen_05}. The injected energy is also thought to control the turbulence in the interstellar medium (ISM) of galaxies \citep{McKee_Ostriker_1977, Elmegreen_2004}.
 
It is challenging to perform numerical simulations of SN explosions to verify whether the physics we have ascribed to them does indeed imply the multitude of observed galaxy properties they are believed to be responsible for, because of the daunting dynamic range between the sub-parsec scale of a single supernova bubble to the $> 10$ kpc scale of the whole galaxy, and because the overlap of SN explosions can dramatically alter the net impact of energy injection \citep{Mori_02}.

Hydrostatic nuclear burning in the massive progenitor stars and explosive nucleosynthesis during the Type ~II SN phase are responsible for producing many of the heavy elements detected in stars and gas \citep[e.g.][]{Woosley_73}, in particular elements containing a multiple number of  Helium ($\alpha$) nuclei.
The accumulation of galactic metals in stars and gas therefore allows a form of \lq archeology\rq, making it possible to infer the history of star formation and SN activity from the metal content of a galaxy. Type II SNe produce r-process elements, enriching the ISM of a galaxy, with some additional help (especially for iron) from type Ia SNe and asymptotic giant branch stars (e.g. \citealp{Burbidge_1957}; see e.g. \citealp{Wiersma_09} for a recent implementation in cosmological simulations). By combining hydrodynamic models of the accretion and outflow of gas from galaxies it should be possible to track the evolution of metals in some detail, as done by e.g. \cite{Pilkington_2012, Brook_2012} where cosmological simulations had individual galaxies re-simulated (zoom simulations) to track the metal distribution in dwarfs. However because of the need to simulate a whole galaxy, these simulations were limited to
a resolution of $25,000\; \Msun$ per resolution element (gas particle), in contrast the simulations described here
better this dramatically to a mass resolution of $0.35 \; \rm \Msun$ per cell. Such high resolution is critical, releasing us from the need to model feedback \lq sub-grid\rq.

Observationally, metals are detected in stellar absorption lines \citep[e.g.][]{Worthey_1994}, nebular lines \citep[e.g.][]{Tremonti_2004}, in galactic winds \citep{Heckman_1990, Heckman_2000, Pettini_2001, Martin_13} and in the intergalactic medium (IGM) \citep{Cowie_1995, Schaye_2003}. Whilst we are far from having a complete inventory of the cosmic metals (e.g. \citealp{Fukugita_98}) due to the selection bias of the tracers and some unobserved sinks such as coronal gas, molecular clouds and low mass stars, we are starting to build constraints on the hydrodynamic processes that transfer metals between the different phases \citep{Finlator_2008, Peeples_2011, Dave_2012}. Because SNe are both the source of energy for driving a galactic wind, and the origin of (some of) the metals observed in winds,  studying the metal distribution of stars, gas and galactic winds could provide valuable information on how SNe drive winds.

Metals enhance the cooling of gas due to their numerous electronic energy transitions, particularly in the range from $10^5 - 10^7 \; \rm K$ \citep{Sutherland_1993,Wiersma_Schaye_and_Smith_09}. In cosmological simulations of galaxy formation, feedback from star formation strives to balance cooling and accretion of gas \citep{Schaye_10}, i.e. when gas cooling is enhanced,  a galaxy will increase its star formation rate and hence the rate at which SNe inject energy to restore equilibrium. Since the presence of metals enhances cooling, obtaining the correct metallicity of halo and ISM is crucial for producing the right amount of star formation in a halo of given mass.

Excitingly, simulations of small galaxies at high redshift are starting to be able to follow in detail how SN explosions stir and enrich the ISM, as well as regulate their star formation by powering a wind, all in a cosmological setting \citep{Wise_12}. Unfortunately it is not yet possible to continue such calculations to the present day, nor investigate how SNe feedback behaves in more massive galaxies, necessitating \lq sub-grid\rq\ models of feedback (e.g. \citealp{Springel_03} and \citealp{Schaye_10}). To bridge the gap between detailed simulations of SNe in a turbulent ISM, in which the calculation resolves the transition of SNe from thermally driven to momentum driven, and such subgrid models, we performed grid simulations
of a patch of galactic disk in \citealp{Creasey_13}, hereafter \CTB. In particular we investigated the relation between the star formation rate, the disk properties, and the strength of the galactic wind. We used the simulation results to make a simple analytic model of how the wind of a galaxy depends on its disk.

In this paper we extend these simulations to follow the highly enriched SN ejecta, using a Chabrier Initial Mass Function (IMF) tracking the metal enrichment and mixing of the different phases, and the subsequent loss of metals into the galactic wind. The paper is organised as follows: in Section \ref{sec:method} we describe the extension to the methodology of \CTB\ to include the tracking of metal production. In Section \ref{sec:betaZResults} we explore the parameter space of disk surface density and potential and define a statistic, $\betaZ$, to encapsulate the metal mass loading. In Section \ref{sec:mass_metallicity} we relate this back to the mass-metallicity relation and then in Section \ref{sec:conclusions} we summarise and conclude.

\section{Methodology}\label{sec:method}
In this section we describe the set of simulations that we use to analyse the metal ejection from galactic disks. As these are an extension of the simulations in \CTB\  we begin with a brief overview of that simulation set up in Section \ref{sec:disk_sims} before describing the addition of SN ejecta in Section \ref{sec:ejecta_method} and their metal composition in \ref{sec:ejecta_comp}. One process absent from \CTB\ was the inclusion of stellar winds, for which we include a prescription in Section \ref{sec:wind_method}, but we do not find that these affect the results greatly.

\subsection{Modelling a supernova-driven galactic wind}\label{sec:disk_sims}
We model a tall column of gas with long ($z$) axis perpendicular to the disk, with the centre of the disk at $z=0$. We assume outflow boundary conditions at the top and bottom of the column, and periodic boundary conditions in $x$ and $y$, and neglect galactic rotation. The equations of hydrodynamics are solved
using \flash3 \citep{Fryxell00}, a parallel, block structured, uniform time-step, adaptive mesh refinement (AMR) code. Integration in space and time are both performed at second order, using a piecewise-parabolic reconstruction in cells. Due to the extremely turbulent nature of the ISM in our simulations, we find that \flash\ attempts to refine (i.e. to use the highest resolution allowed) almost everywhere within our simulation volume. Therefore we disable the \amr\  capability of \flash\ and run it at a constant refinement level, i.e. using a fixed grid. To mitigate the overhead of the guard-cell calculations we increase our block size to $32^3$ cells per block.

The gas is assumed to be mono-atomic with equation of state
\begin{equation}
p = (\gamma-1)\rho\,u\,,
\end{equation}
where $\gamma=5/3$. For our primary set of simulations gas cools at a rate dictated by a cooling function $\Lambda$ that depends on temperature as
\begin{equation}\label{eq:cooleq}
\rho \dot{u} = \left\{ \begin{array}{cc} 
 -\Lambda n^2 ,&  T \geq T_0 \\
0 , & T< T_0\,,  \end{array} \right.
\label{eq:cf}
\end{equation}
where we in addition assume pure hydrogen gas so that the number density $n = \rho / m_p$, and $\Lambda=10^{-22} \, \rm erg \, cm^3 \, s^{-1}$, however we also investigate the effect of a more realistic (and metal dependent) cooling function.
Cooling is truncated below $T_0 \equiv 10^4 \; \rm K$ to prevent collapse into molecular clouds (which would be unresolved in our simulations). Consequently we only follow the warm ($\sim 10^4 \; \rm K$) and hot ($\sim 10^6 \; \rm K$) phases of the ISM, which form a bi-modal distribution due to the fast cooling time of gas at intermediate temperatures in pressure equilibrium. 

In \CTB\ we explored simulations which assumed the more realistic cooling function of \cite{Sutherland_1993} but found our results to be surprisingly insensitive to the detailed temperature dependence of $\Lambda$. Essentially, when cooling was efficient, the cooling time was so short that its detailed dependence on temperature is mostly irrelevant for the dynamics of the simulations, i.e. the normalisation of the cooling function was important for the transition between the phases but not the phases themselves. Our motivation for repeating such a simplified cooling function is to be able to distinguish features in the temperature distribution of the gas that are due to cooling, versus those that are a consequence of SN heating and gas outflow, however in this work there may be an additional complication due to the inhomogeneous metal distribution, and as such we also perform several simulations with the metal dependent cooling rates. 

We model gravity of the disk due to stars, gas, and dark matter, by imposing a time-independent gravitational potential $\phi$, which is the solution of 
\begin{equation}
\nabla^2 \phi = 4\pi G {\rho\over \fg}\,,
\end{equation}
where $\rho$ is the gas density at the start of the simulation and $\fg$ is the initial gas fraction, calculated for each grid cell. We note that the time independent approximation is only valid as long as both the mass of the simulation remains close to the initial mass and that no structures develop with mass above the Jeans mass (at temperature $T_0$). Since the simulation begins in hydrostatic equilibrium at $T=T_0$ the Jeans mass is the that of the disk, and thus the latter condition is automatically satisfied. For the former condition we require that the simulations are performed for a sufficiently short time that the mass does not evolve, and we have explicitly verified that the relative mass loss in our simulations is $(0.42^{+0.29}_{-0.19})\%$, over the time frame considered. We note that for the real ISM the peaks in the gas distribution reach sufficient densities that molecular cooling becomes important, taking them to even lower temperatures and smaller Jeans masses (i.e. they are molecular clouds), so self-gravity does become important again on these scales. Unfortunately we do not resolve such small scales in our simulations, indeed even with a cooling function for the cold gas it seems that simulations at this resolution are unable to cool to molecular temperatures \citep[see e.g.][]{Gent_2012}. The small scale density structures in our simulations are thus driven by cooling and turbulence rather than gravity.

We further assume the disk makes stars at a rate set by the Kennicutt-Schmidt (KS, \cite{Kennicutt_98}) relation, for which the surface density rate of star formation is related to the gas surface density by
\begin{equation}
\dot{\Sigma}_\star =  2.5\times 10^{-4} \left({\sigmag\over \Msun\,{\rm pc}^{-2}}\right)^{1.4} \, \rm \Msun \, yr^{-1} \, kpc^{-2} \,.
\label{eq:KS}
\end{equation}
As the simulation progresses in time, we keep the surface density rate of star formation equal to its initial value -- and hence also the imposed start formation rate remains constant in time. This is a valid approximation since our simulations are evolved over a time much shorter than the gas depletion time scale.

We translate the star formation rate into a core-collapse supernova rate assuming each 100~$\Msun$ of stars formed yields $\epsilon_{100}$ SNe\footnote{For reference, for a \cite{Chabrier_03} stellar initial mass function composed of stars with masses between 0.1 and 100~$\Msun$, $\epsilon_{100}=1.8$ assuming those in the range $[6,100]~\Msun$ undergo core collapse.}.
We further assume that the star formation rate is proportional to the local initial density, i.e. the probability per unit volume that a SN explodes is proportional to both the star formation surface density, and the gas density\footnote{We do not assume that SNe explode at density peaks, even though the progenitor stars may have formed there.
Our motivation for this is that the star may have drifted out of it natal molecular cloud - or that that cloud has been destroyed by say by radiation - in the $~\sim 30$~Myr between the formation of the star and its final explosion as a SN.},
\begin{equation}
\mathbb{E} [ \dot{\rho}_\star {\rm \, dV \, dt} ] = \dot{\Sigma}_\star \frac{\rho(t=0) }{\sigmag} {\rm \, dV \, dt}\, .
\end{equation} 
Simulations similar in spirit have been performed by \citet{Slyz_2005, Joung_MacLow_2006, Hill_2012} and \citet{Gent_2012}. 
In \CTB\ we modelled a SN explosion by injecting $10^{51}$~ergs of thermal energy at the location of the explosion. In the current paper we have improved on this as described below. As initial conditions we assume that the gas is in hydrostatic equilibrium, with uniform temperature $T=T_0\equiv 10^4$~K.

In \cite{Creasey_13} we showed that the action of SNe changes the initial gas distribution on a short time-scale of $\sim$~$10$~Myrs into a turbulent ISM with large range in temperature and density, in order-of-magnitude pressure equilibrium. Occasionally several SNe will explode by chance in close proximity in rapid succession, leading to a locally significantly over-pressurised ISM. Such high pressure bubbles expand away from the galactic plane accelerating as they convert thermal energy into kinetic energy. As a result, expansion speed increases proportional to height above the plane, resembling a rarefaction wave. The \lq galactic wind\rq\ is the time-averaged result of many such waves. As the rarefaction wave punches through the disk it envelopes some of the cooler disk material dragging it along, increasing the mass flux of the wind above of that of just gas heated by the SNe. We characterised this \lq mass loading\rq\ by
\begin{equation}
\beta \equiv {\dot M_{\rm w}\over \dot M_\star}\,,
\label{eq:beta}
\end{equation}
i..e the winds mass flux, $\dot M_{\rm w}$, measured by the flux of material escaping the computational column through its top and bottom face, in units of the star formation rate. Running a suite of simulations in which we varied $\Sigma$ and $\fg$, \cite{Creasey_13} determined that the mass loading depends on total surface density $\Sigma$, gas surface density $\sigmag$ and gas fraction $\fg$ approximately as

\begin{eqnarray}
\label{eq:beta_fit}
\beta &=& (0.63\pm 0.49)  \left({\sigmag\over 10 \Msun\,{\rm pc}^{-2}}\right)^{-1.15 \pm 0.12} \left({\fg\over 0.1}\right)^{0.16\pm 0.14}\nonumber\\
&=& (9.9\pm 7.7)  \left({\Sigma\over 10 \Msun\,{\rm pc}^{-2}}\right)^{-1.15 \pm 0.12} \left({\fg\over 0.1}\right)^{-1\pm 0.14}\,.
\end{eqnarray}
where in the latter we have alternatively parameterised in terms of the total (gas and stellar) surface density $\Sigma$ rather than the gas surface density $\sigmag$ that we use in Section \ref{sec:betaZResults}.

\subsection{Supernova ejecta}\label{sec:ejecta_method}
In  \CTB\  we modelled SNe explosions simply by increasing the pressure in the vicinity of the explosion by
\begin{equation}
\Delta p(\mathbf{r} ) = (\gamma -1)\frac{E_{\rm SN}}{\left(2 \pi r_s^2 \right)^{3/2}} \exp \left(-\frac{1}{2} \frac{r^2}{r_s^2}\right) \, .
\end{equation}
Here, $E_{\rm SN}=10^{51} \; \rm erg$ is the bulk kinetic energy released by a single SN, thermalised by shocking into the ISM, and $r$ is the distance from the centre of the SN. The sharpness of the pressure spike is set by $r_s$, which we took to be $r_s=2$~pc, a compromise between resolution (i.e. not smearing the supernova over too large a volume) and allowing for the limitations of the hydrodynamics solver (by not placing the energy in just 1 cell). Given the numerical resolution of the simulations, this width is so small that the SN is still in the energy driven phase, provided the density $n<77$~cm$^{-3}$ ($\rho<1.3\times 10^{22}$~g~cm$^{-3}$, see also \CTB), so that the simulations model the transition from a thermally driven shell (Sedov-Taylor explosion) to a momentum driven shell (\lq snow plough\rq\ phase). 

We have improved the modelling of SNe by including SNe ejecta, which we treat as an additional fluid, $\rho = \rho_p + \rho_e$, where $\rho_p$ is the original (pristine) gas and $\rho_e$ is the gas injected by SNe. Initially $\rho_e$ is set to zero everywhere and is increased by each SN by
\begin{equation}
\Delta \rho_e(\mathbf{r} ) = \frac{M_{\rm SN}}{\left(2 \pi r_s^2 \right)^{3/2}} \exp \left(-\frac{1}{2} \frac{r^2}{r_s^2}\right) \, ,
\end{equation}
where $M_{\rm SN}=10 \; \rm \Msun$ is the total mass added. This value corresponds to a progenitor stellar mass of $\approx 12 \; \rm \Msun$ \citep{Woosley_1995}, chosen to be representative of a Chabrier IMF where core collapse occurs for stars in the range $[6, 100] \; \rm \Msun$.  

Notably this sets a (maximum) specific energy for the remnant which was absent from the simulations of \CTB. In those simulations the SNe could explode in arbitrarily sparse environments, and there would be a (very small) tail of gas at $T > 10^9 \; \rm K$. In the current simulations the SNe energy inject gas at a maximum temperature of
\begin{eqnarray}
T_e &\equiv& \frac{\mu m_{\rm H}\Delta p}{k_{\rm B} \Delta \rho_e} \nonumber \\  
 &=& (\gamma - 1) \frac{\mu m_{\rm H}\,E_{\rm SN}}{k_{\rm B} M_{\rm SN}}  \nonumber \\
&\approx & 2.4 \times 10^8 \; \rm K\, {E_{\rm SN}\over 10^{51}{\rm erg}}\,\left({M_{\rm SN}\over 10\,\Msun}\right)^{-1}\,.
\label{eq:ejec_temp}
\end{eqnarray}
The \flash\ code will advect the ejecta along with the original gas allowing us to examine enrichment.

\subsection{Cooling and the metal composition of ejecta}\label{sec:ejecta_comp}
The specific choice of $10 \; \rm \Msun$ for the mass of the SN ejecta may at first appear crucial to the subsequent metallicity evolution of the ISM, as this sets the quantity of metals that are introduced per event. The hydrodynamics of our simulations, however, are only weakly dependent on this mass because the evolution of the remnant is primarily driven by the energy of the SN, via the temperature in Eq. (\ref{eq:ejec_temp}). Therefore to reduce the number of simulations needed 
we keep our ejecta mass fixed and simply assume that their metallicity varies with the yield, i.e.

\begin{eqnarray}
\Zej &\equiv & \frac{M_{Z,\rm SN}}{M_{\rm SN}} \label{eq:Z_ej}\\
&=& y  \left( \frac{100 \epsilon_{100} \rm \Msun}{M_{\rm SN}}\right) \, , \nonumber
\end{eqnarray}
where to be consistent with \CTB\  we use $\epsilon_{100} = 1$ SN per $100 \; \rm \Msun$ of stars formed, and the yield $y$ refers to the mass of oxygen released into the ISM per $1 \; \rm \Msun$ of star formation. This analysis could be repeated for other elements. However, for other elements (particularly iron) the departure from instantaneous recycling due to the importance of long lived stars on the returned fraction will make the approximations progressively poorer (e.g. \citealp{Schmidt_1963, Tinsley_1980}). 

We will take as fiducial value a yield of $y=0.02$. Theoretically $y$ is probably only known to within a factor of $\sim 2$ (\citealp{Woosley_1995, Finlator_2008}, hereafter FD08). Therefore where possible we quote in fractions of $y$ to reduce this uncertainty. In order to retain the scale-free nature of the calculation this means that we must also choose a cooling function that is independent of metallicity, and for our primary cooling function we continue using the function of \CTB\ (Eq.~\ref{eq:cf}). This makes the specific value of the solar metallicity $Z_\odot$ is only of interest for reference purposes in these calculations, and we used the value $Z_\odot=0.0165$ \citep{Asplund_2005}. Given the values in the previous paragraphs this is around $8\%$ of the ejecta metallicity (i.e. the Sun has formed from diluted material). We are also making the assumption of chemically identical type II SN, in reality the ejecta of progenitor stars that differ in initial mass will have different abundance patterns.

In order to explore the sensitivity to metallicity, however, we additionally include a set of simulations with the metal-dependent cooling function of \citet{Sutherland_1993}, which enhances the rate of cooling around $10^6 \; \rm K$ (see e.g. \citealp{Wiersma_Schaye_and_Smith_09}). This breaks the scale-free nature of our simulations, i.e. the initial metallicity of simulation volume becomes important. To bound the parameter space we have included simulations where the initial metallicity is uniform, with values $0$ (pristine), $0.1\, Z_\odot$ and $Z_\odot$. Over the short time scales ($20$ Myr) of these simulations, an ISM at solar metallicity contains a significantly higher density of metals than will be injected via SNe.

\subsection{Stellar winds}\label{sec:wind_method}
\begin{figure*}
\centering
\includegraphics[width=2\columnwidth]{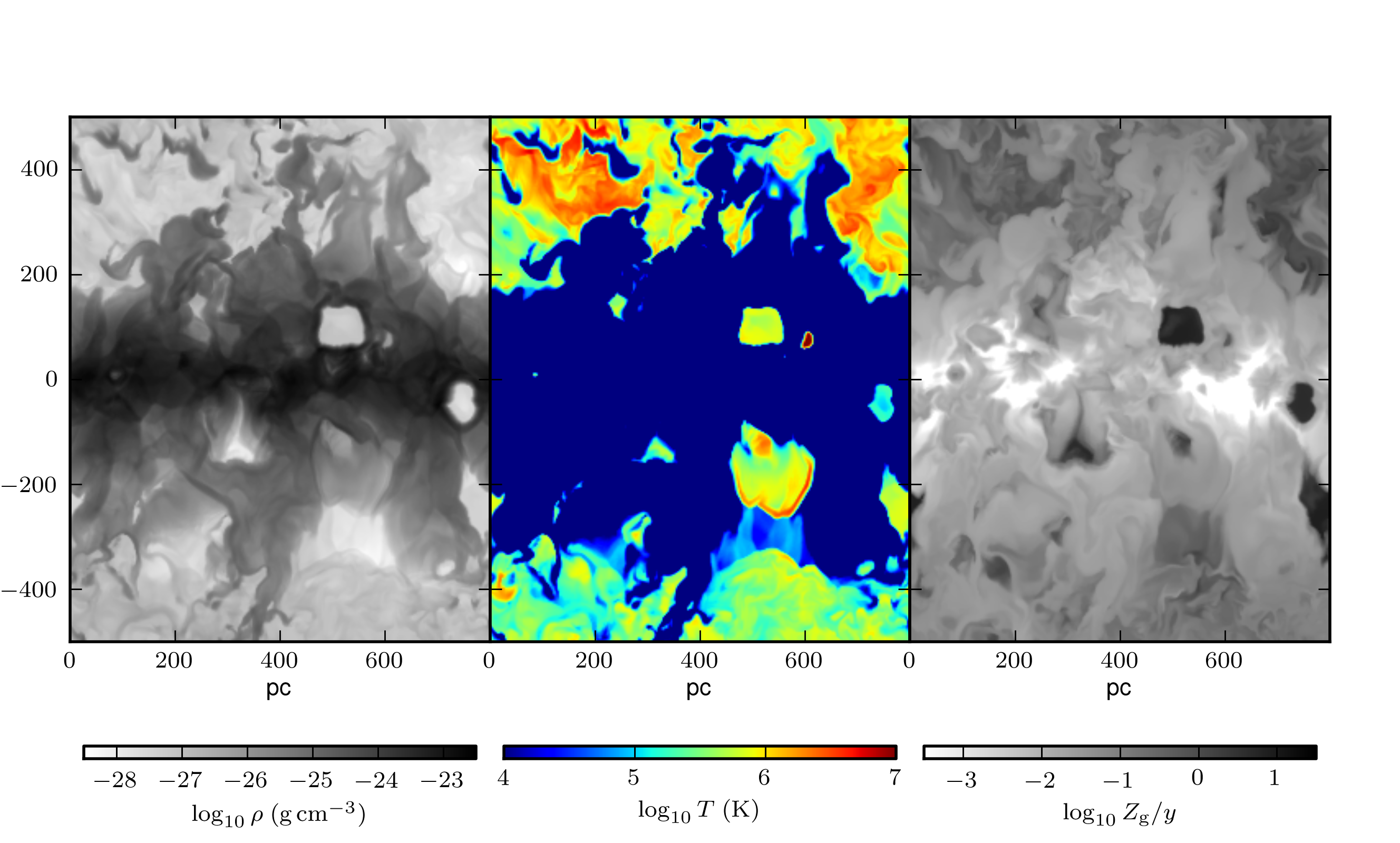}
\caption[Metal ejection from a disk showing density, temperature and metallicity]{Slice in the $x-z$ direction through the simulation volume at 15 Myr perpendicular to the disk; the disk mid-plane is at $z=0$. Panels from {\em left} to {\em right}, depict gas density, temperature, and metallicity, respectively. The disk is stirred by many generations of SNe but remains mostly intact: it is the cold and dense band of gas near $z=0$. The disk gas is punctured by young SNe remnants, which appears as low-density, hot, and highly enriched bubbles. Previous generations of SNe have launched a wind, which can be seen as the tenuous enriched gas at a range of temperatures undergoing turbulent mixing, out-flowing both below and above of the disk.}
\label{fig:pic_metals}
\end{figure*}

\begin{figure*}
\centering
\includegraphics[width=2\columnwidth]{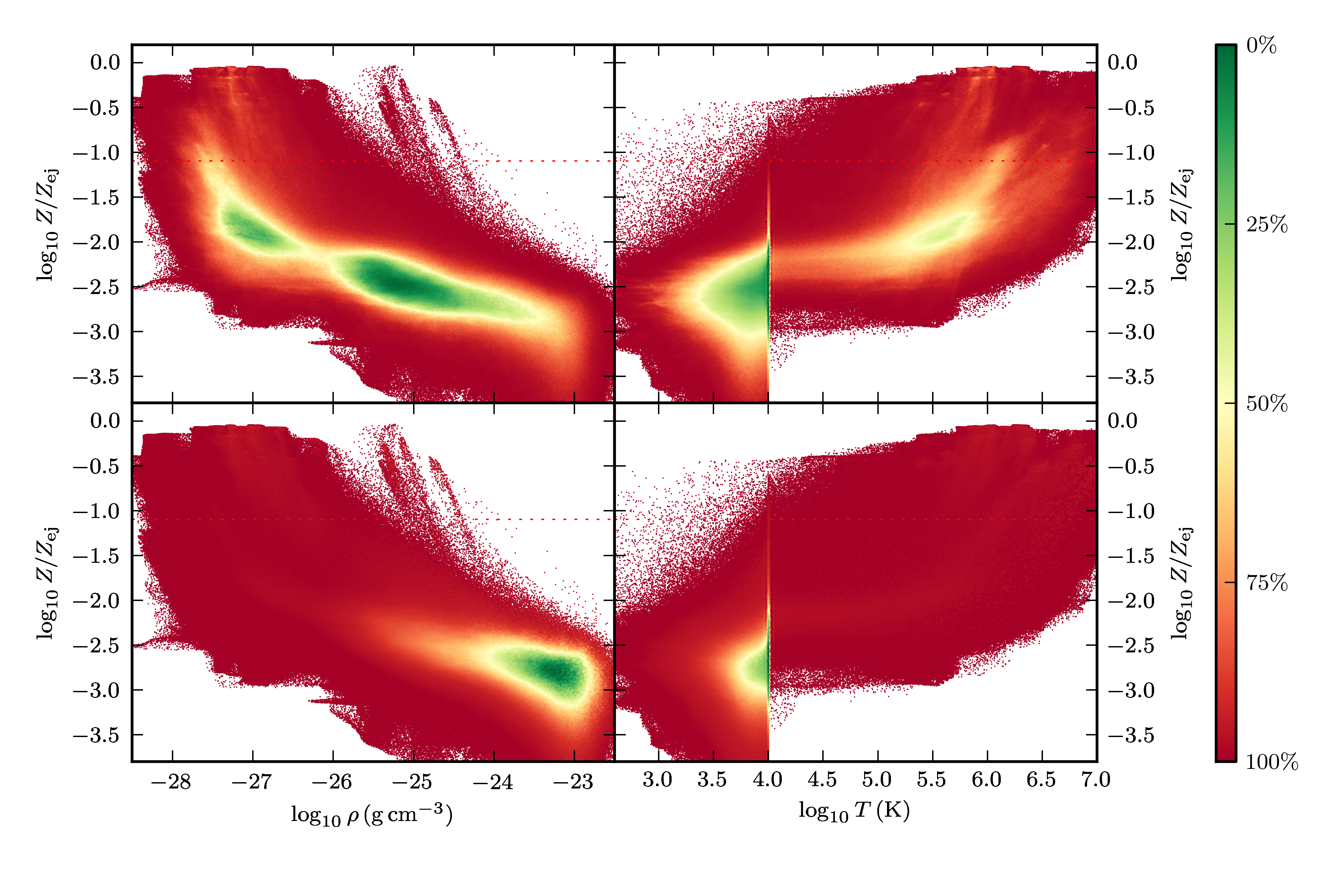}
\caption[Metallicity-density and metallicity-temperature phase space diagrams.]{Density vs. metallicity ({\em left panels}) and temperature vs. metallicity ({\em right panels}) phase space diagrams of the ISM for the simulation shown in Fig.~1;
{\em top panels} are shaded by volume enclosed by each phase (percentage enclosed is indicated by the colour bar), whilst the \emph{lower panels} are shaded by the fraction of metals in each phase. The vertical feature at $10^4 \; \rm K$ is the base of the cooling function, and the \emph{red dashed line} indicates solar metallicity assuming a yield of $y=0.02$ (see Eq. \ref{eq:Z_ej} for details). Most of the metals lie in the warm dense phase ($T \sim 10^4\;\rm K$, $\rho \sim 10^{-26}$~-~$10^{-23} \; \rm g \, cm^{-3}$), but the hot phase has a higher average metallicity; with abundance increasing smoothly with decreasing density or increasing temperature.  The scatter in metallicity is large, $\sim 1$~dex, in all phases.  The semblance of a reflection symmetry between the \emph{left} and \emph{right panels} is due to the (approximate) pressure equilibrium of the ISM, i.e. density is the inverse of temperature, within an order of magnitude. Gas below $T=10^4~K$ arises due to adiabatic expansion. }
\label{fig:Z_phase}
\end{figure*}

We ran a number of simulations that include the winds from massive progenitor stars, in addition to the energy injected by SNe. We consider these to occur at the same sites as the SNe, and act uniformly over a time of 10 million years. They can thus be considered either to issue from the progenitor stars of the SNe or of the OB associations in which the SNe occur.

The prescription for injecting stellar winds is the introduction of thermal and kinetic energy, in a similar way to the instantaneous energy injection for the SNe. The mass, momentum, and energy source due to the precursor of a SN are
\begin{eqnarray}
\left. \dot{\rho} \right|_{\rm OB} &=& 
\frac{\dot{M}_{\rm OB}} {(2 \pi r_s^2)^{3/2}}
\exp \left(-\frac{1}{2} \frac{r^2}{r_s^2}\right) \\
\left. \dot{p} \right|_{\rm OB} &=& (\gamma - 1) 
\frac{\dot{E}_{\rm OB}} {(2 \pi r_s^2)^{-3/2}} \exp \left(-\frac{1}{2} \frac{r^2}{r_s^2}\right) \\
\left. \frac{\partial}{\partial t} \left( { \rho \bf v}\right) \right|_{\rm OB} &=&  \frac{\bf r}{r_s} \frac{\dot{P}_{\rm OB}}{8 \pi r_s^3} \exp \left(-\frac{1}{2} \frac{r^2}{r_s^2}\right) \, ,
\end{eqnarray}
where we have normalised the rates such that the rate of mass injection is $\dot{M}_{\rm OB}$, the rate of thermal energy injection $\dot{E}_{\rm OB}$, and the rate of (absolute) momentum injection $\dot{P}_{\rm OB}$. Our fiducial values for these rates are
\begin{eqnarray}
\dot{M}_{\rm OB} &=& 0.1 \; \rm \Msun \, Myr^{-1} \\
\dot{E}_{\rm OB} &=& 10^{50} \; \rm erg \, Myr^{-1} \\
\dot{P}_{\rm OB} &=& 920 \; \rm km \, s^{-1} \, \Msun \, Myr^{-1}  \, ,
\end{eqnarray}
where the momentum injection corresponds to a kinetic energy injection rate of approximately $\frac{1}{2} \dot{P}_{\rm OB}^2 \dot{M}_{\rm OB}^{-1} \approx 8.5 \times 10^{49} \; \rm erg \, Myr^{-1}$ (close to that of \citealp{Castor_1975a}), though this will to some extent depend on the local environment. Over the 10 million years
over which the wind is assumed to act, it will have released an additional $E_{\rm SN}$ in thermal energy and almost the same again in mechanical energy. We have intentionally not chosen conservative values for these energy injection rates in order to make the effect(s) of winds more apparent in our simulations. The high implied wind velocity ($\dot{P}_{\rm OB} / \dot{M}_{\rm OB}$) is due to the addition of the radiation driving of the winds (see also \citealp{Murray_2005}), i.e. the massive stars release a large fraction of energy as radiation which couples to more than just the mass in the wind (e.g. \citealp{Hopkins_2011a}), avoiding a full radiative transfer calculation on the mesh (although such calculations are becoming possible, see for example \citealp{Rosdahl_2013}).

\section{Metal mixing and the rate of metal ejection}\label{sec:betaZResults}
In this section we describe the evolution of metallicity in our simulations. We present the distribution of metallicity within the ISM in terms of the different phases and discuss the effects of our parameter choices. We then move on to looking at the out-flowing metals that escape from the galactic disk and the dependency of this on the disk properties. Finally, we discuss the origin of the correlation between thermalisation and metal mass loading from the disk. We will refer to all the gas in the computational volume (height $|z|<500$~pc) as the \lq interstellar medium\rq, whereas the gas that exits the volume through the top and bottom faces of the column as the \lq wind\rq. This definition of the interstellar medium approximately covers the volume that encloses the neutral material in our own Milky Way (for an overview of disk components see, for example, \citealp{Holmberg_2004}).
The majority of the gas by mass at low $|z|$ remains warm ($T\sim 10^4$~K) and dense ($\rho\sim 10^{-24-23}$~g~cm$^{-3}$), and we will refer to this as the \lq disk\rq; above the disk is the launch region of the wind.

\subsection{The metallicity of the ISM}\label{sec:ism_metallicity}
We begin with an illustrative slice through a simulation volume of width 800~pc at a time\footnote{We showed in \cite{Creasey_13} that these simulations enter a quasi-steady state within (at most) 10~Myrs, therefore the choice of time subsequent to this is relatively unimportant (ignoring the exact spatial location of individual turbulent features).}  of 15 Myr for a disk with gas surface density of $\sigmag=11.6 \; \rm \Msun \, pc^{-2}$ and gas fraction $\fg = 0.1$, see Fig.~\ref{fig:pic_metals}. At the mid-plane, $z=0$, we see the disrupted disk, where the initially pristine gas has been stirred and enriched by several generations of SNe. In a few regions individual recent SN remnants are discernible, where they stand out as noticeably hot and sparse highly enriched bubbles. Above and below the disk is the launch region of the turbulent hot wind (see \CTB) with higher mean metallicity than the disk. Comparing the panels reveals a correlation between temperature and metallicity, and an anti-correlation between density and metallicity, which we will discuss further in Section \ref{sec:metal_diffuse}.

In Fig.~\ref{fig:Z_phase} we take a closer look at the ISM by studying the phase space in metallicity vs. density and temperature at a snapshot of the fiducial simulation at 10 Myr.
The hot, sparse phase ($T>10^5$~K, $\rho<10^{-26}$~g~cm$^{-3}$) has a significantly higher median metallicity than the warm dense phase, reaching values close to 10 times solar, with a very large scatter. Its mass-weighted metallicity is higher than the volume weighted one, meaning that metals tend to be locked-up in over-dense clouds. In terms of total mass in metals, the warm phase dominates due to its much greater mass of gas. Our reference value of $8\%$ for solar metallicity as a fraction of the ejecta metallicity lies at the lower range of metallicities for the hot phase but is significantly higher than the metallicities of the warm phase. Of course the metallicity of the warm phase is not static, as it is being steadily enriched by successive generations of SN, and after several giga-years we would expect the cooler material to be sufficiently enriched to form higher metallicity stars. The gas with temperature below $10^4 \; \rm K$ is due to adiabatic expansion, in the turbulent ISM the compression and expansion combined with cooling in the compressed phase allows the gas to scatter below $10^4 \; \rm K$.

\begin{figure*}
\centering
\includegraphics[width=2\columnwidth]{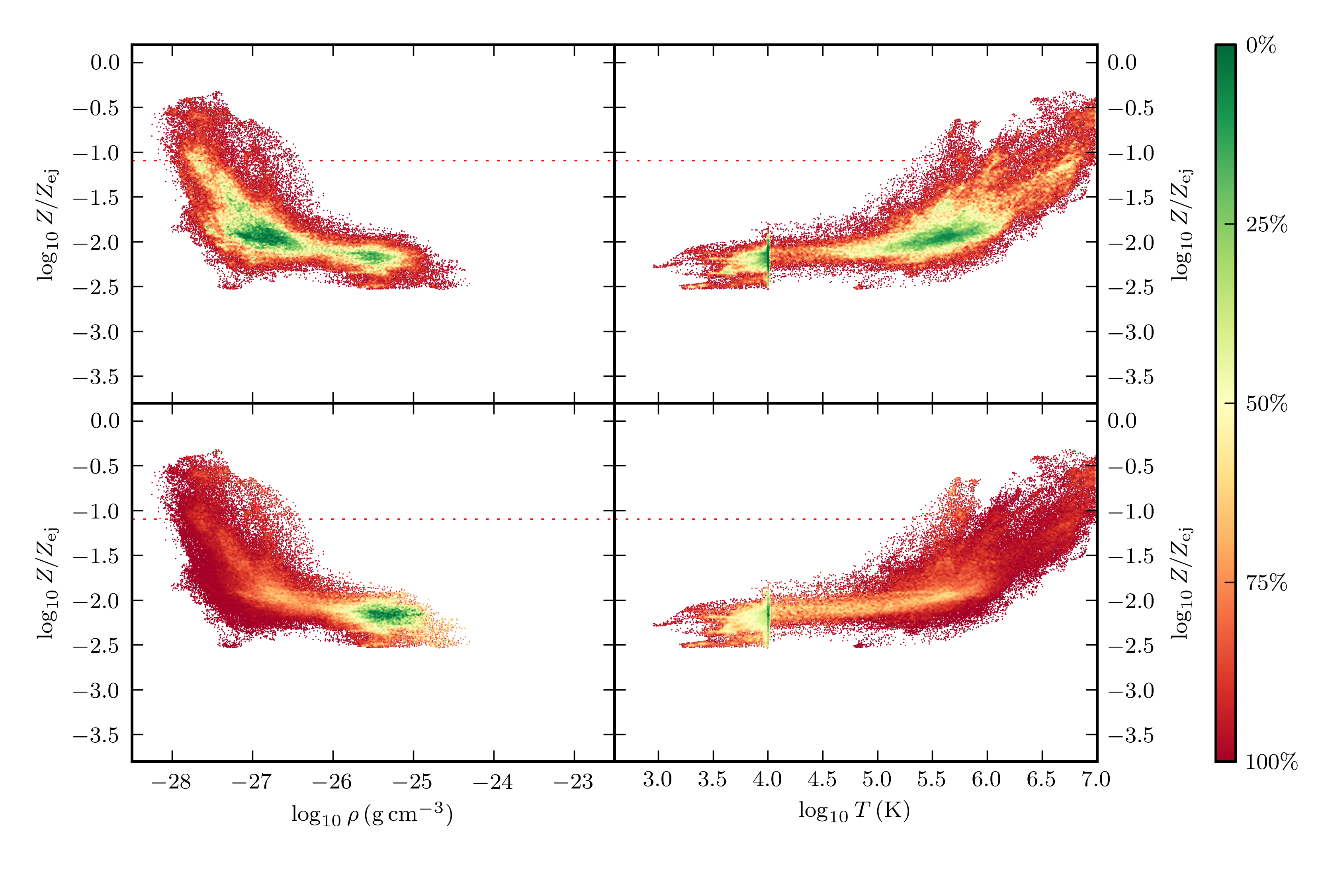}
\caption[As for Fig.~\ref{fig:Z_phase} but only for the material being ejected from the disk.]{As for Fig.~\ref{fig:Z_phase} but only for the material being ejected from the disk, and out-flowing the computational volume at height $|z|>500$~pc. Warm  dense gas ($10^4$~K, $\rho\sim 10^{-26-23} \, \rm g \, cm^{-3}$) is entrained by the hot  tenuous ($10^{5-7}$~K, $\sim 10^{-26-28} \, \rm g \, cm^{-3}$) outflowing wind, and has a metallicity typically 1~dex lower. }
\label{fig:Z_phase_edge}
\end{figure*}

In contrast to the ISM, the metallicity of the unbound hot phase is approximately time independent, as the injected metals can escape from the disk and that gas is not continually
enriched by successive generations of stars\footnote{Some of this gas may cool and fall-back onto the disk, to be expelled once more, but this cannot occur in our simulations as they have outflow boundary conditions.}.  In Fig.~\ref{fig:Z_phase_edge} we restrict our attention to the outflowing material, that leaves the computational volume at the top or bottom of the simulated column. This gas is dominated by the hot, low density, high metallicity phase, indeed there is no gas with density above $10^{-24} \; \rm g \, cm^{-3}$ or relative metallicity below $\approx 10^{-2.5} \, \Zej$. Denser gas is entrained ISM, which is at lower metallicity and is considerably cooler. 
As discussed in \cite{Creasey_13}, such entrainment of dense gas in the wind is crucial to obtain the high values of the mass loading, $\beta$ given in Eq.~(\ref{eq:beta}), required by models of galaxy formation. The nature of our enrichment mechanism entails that the outflowing gas from galactic disks will be mostly of significantly higher metallicity than the average gas phase metallicities of the ISM, because it is the hot SN material that drives the wind. Unfortunately, as this gas is rather hot and of low column density it is hard to observe directly, and it may be easier to derive constraints from the X-ray coronae of halos (e.g. \citealp{Crain_2010}). There is also the complication that this gas will quickly mix with material in the circum-galactic medium (CGM) to form a lower metallicity blend.

\subsection{Dependence on disk parameters}

\begin{figure}
\centering
\includegraphics[width=\figwidth]{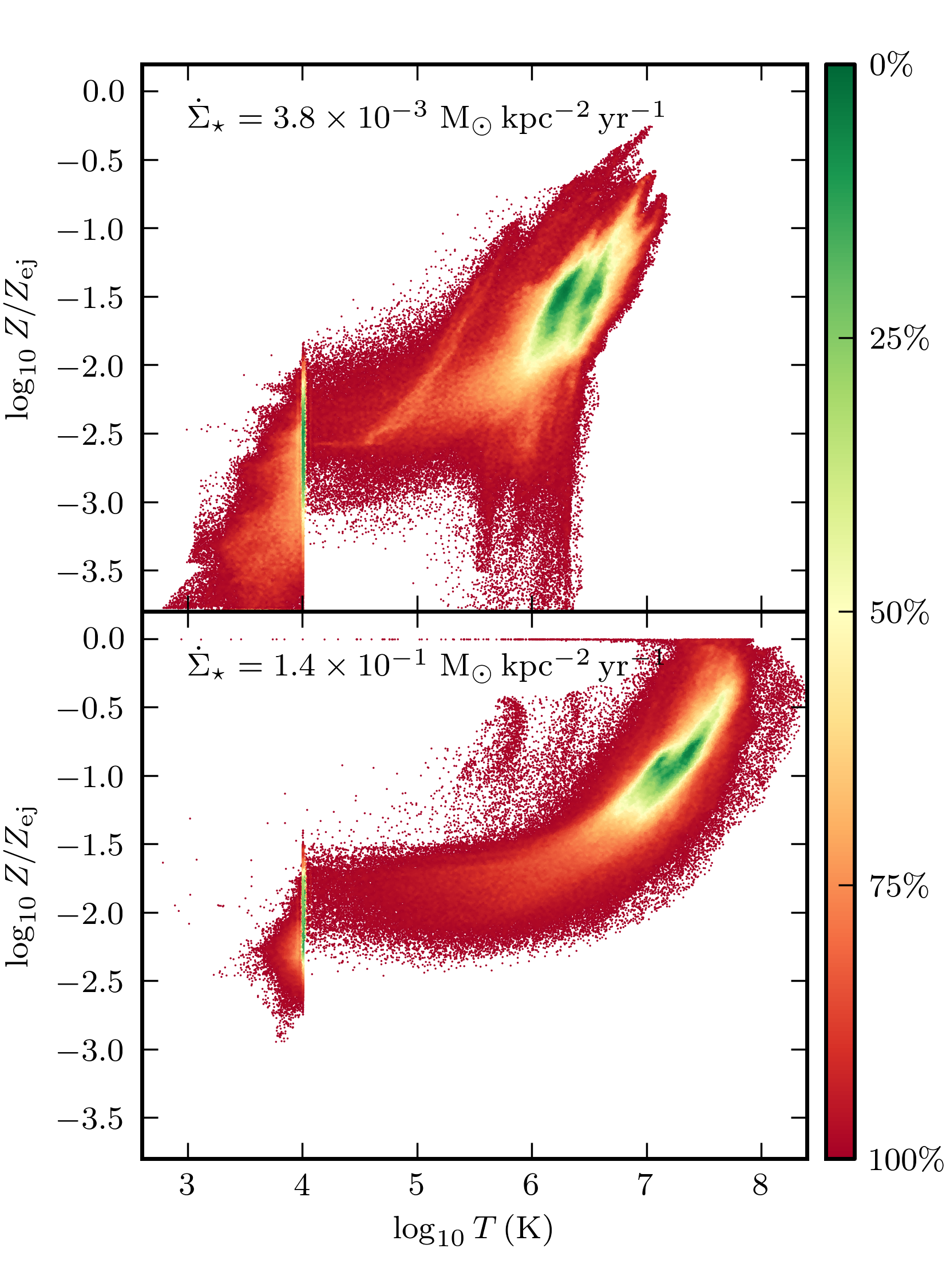}
\caption[Comparison of ISM temperature-metallicity phases for disks with different star formation rates.]{Dependence of the ISM temperature-metallicity phases on the gas surface density of the disk, comparing disks with gas and star formation surface densities of
$\sigmag=92~\Msun$~pc$^{-2}$, $\dot\Sigma_\star=1.4\times 10^{-1}\,\rm \Msun \, kpc^{-2} \, yr^{-1}$ ({\em bottom panel}) to a disk with considerably lower gas density and star formation rate,
$\sigmag=7~\Msun$~pc$^{-2}$, $\dot\Sigma_\star=3.8\times 10^{-3}\,\rm \Msun \, kpc^{-2} \, yr^{-1}$ ({\em top panel}). The bottom disk with the higher star formation rate has a considerably hotter outflow, which is also considerably more metal enriched.}
\label{fig:two_ism}
\end{figure}
\begin{figure}
\centering
\includegraphics[width=\figwidth]{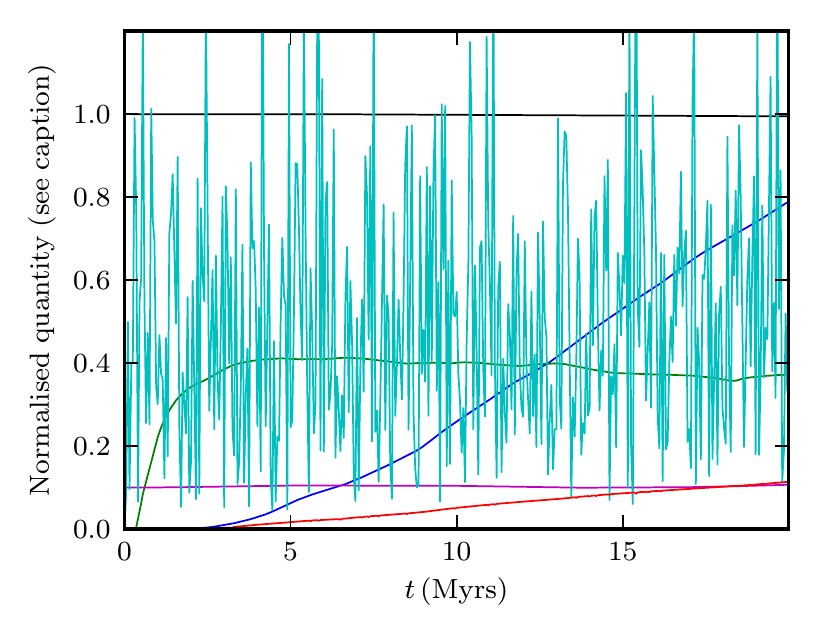}
\caption[Time evolution of a metal ejection simulation]{The time evolution of the simulation shown in Fig.~\ref{fig:pic_metals}. \emph{Cyan line} is the cooling rate of the simulation volume as a fraction of the mean SN heating rate, \emph{black line} is the fraction of gas remaining in the simulation volume, \emph{green line} is the porosity normalised as $0.2P+0.5$.; \emph{dark blue line} is the surface density of gas ejected from the disk in units of $0.1\; \rm \Msun \, pc^{-2}$, \emph{red line} is the ejected surface density of metals, in units of $0.002\; \rm \Msun \, pc^{-2}$. Finally the \emph{magenta line} is the disk's scale height, in units of its initial value divided by 10.  The cooling rate fluctuates wildly, with peaks associated with SN events. After a short time, $t\sim 2$~Myrs, the porosity $P$ reaches a steady state, and shortly afterwards ($t\sim 5$~Myrs) a steady outflow of mass and metals starts up. The total amount of mass in the simulation, and the disk's scale height, remain nearly constant.}
\label{fig:ej_time}
\end{figure}

In \CTB\ we demonstrated that higher gas surface density disks, that, according to the Kennicutt-Schmidt law of Eq.~(\ref{eq:KS}), have higher star formation rates, have a higher temperature outflow, and consequently lower mass loading, $\beta\propto \sigmag^{-1.15\pm 0.12}$, primarily because such denser disks have both a stronger gravitational potential, and a higher cooling rate. Consequently,
to escape from this disk the outflow needs to be both hotter and at lower density. We expect this to carry through to these simulations that include the ejecta, and also that there will be a corresponding trend in metallicity, where the outflows in simulations with higher surface densities entrain less of the surrounding gas and are therefore both hotter and more metal rich. In Fig.~\ref{fig:two_ism} we probe this by comparing the ISM at a given time for two different gas surface density disks. We can indeed see that the peak of the distribution of the hot phase for the higher surface density simulation lies at a temperature and metallicity nearly an order of magnitude higher than that for the lower surface density simulation.

We now turn our attention to the time evolution of the ISM. In \CTB\ we saw how the stochastic mass ejection from the disks could be averaged over a large number of events to estimate a mean outflow rate for a given idealised disk. We attempted to measure this by combining the net mass loss from several snapshots in time and performing a linear fit with some delay time, such that the slope of the mass loss would indicate the outflow rate. In Section \ref{sec:met_outflow_deps} we will attempt the corresponding analysis, but in terms of the metal ejection rate, and for this reason we investigate the time evolution for a single simulation.

Using normalised quantities, we plot in Fig.~ \ref{fig:ej_time} the time evolution of
\begin{itemize}
\item[({\em i})] the total radiative cooling rate in units of the mean injection rate of SN energy, i.e. the volume integral of Eqn.~(\ref{eq:cooleq}), divided by the energy injection rate of SNe, $\epsilon_{100} E_{\rm SN} \dot{M}_\star / (100 \Msun )$,
\item[({\em ii})] the fraction of total gas mass remaining in the computational volume
\item[({\em iii})] the porosity, $P\equiv -\log(f_{\rm warm})$, where $f_{\rm warm}$ is the volume fraction of warm gas ($T<2\times 10^4$~K),
\item[\emph{(iv)}] the surface density of gas ejected and 
\item[\emph{(v)}] the surface density of metals ejected from the simulation volume 
\end{itemize}
for the simulation in Fig.~\ref{fig:pic_metals}.

 As noted in Paper I, the normalised cooling rate fluctuates wildly and is highly correlated with the individual SN events. Shortly after each SN event ($\sim 10,000$ yrs) the blast wave ceases to be adiabatic and suffers heavy radiative losses as it runs into dense patches of gas. The normalised cooling rate can reach values higher than the mean energy injection rate following some SN episodes, although the median value is much lower at $\sim 0.4$. After a period of $\approx 2 \; \rm Myr$ the porosity of the ISM has converged to a value of $P\sim -0.5$ ($f_{\rm warm}\sim 0.3$), and after $\sim 5$~Myrs, an outflow of mass and metals starts, with mass loss and metal mass loss proceeding at a nearly constant rate. Over the simulation time the total mass of gas and the scale height of the disk do not evolve appreciably.

\begin{figure}
\centering
\includegraphics[width=\figwidth]{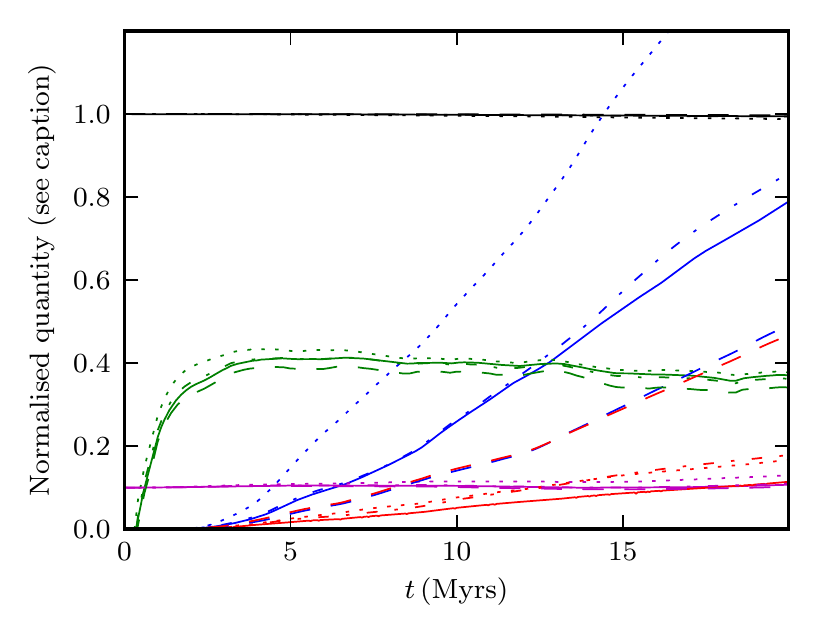}
\caption[Time evolution with metal-dependent cooling]{The time evolution of the porosity (\emph{green line}), surface density ejection (\emph{dark blue line}), metal surface density ejection (\emph{red line}), disk scale height (\emph{magenta}) and mass remaining (\emph{black line}) using the normalisations from Fig.~\ref{fig:ej_time} (we exclude instantaneous cooling for clarity) for the same simulation, along with simulations using with the metal-dependent cooling function of \citet{Sutherland_1993} with 3 different initial gas metallicities. \emph{Dotted line} is initially pristine zero-metallicity gas, \emph{dot-dashed line} is $0.1 Z_\odot$ and \emph{dashed line} is $Z_\odot$.}
\label{fig:metaldep_cool}
\end{figure}

In order to understand the effects of metal-dependent cooling we compare this fiducial simulation to three simulations with a metal-dependent cooling function \citep{Sutherland_1993} in Fig.~\ref{fig:metaldep_cool}, using initial metallicities $0$, $0.1 Z_\odot$ and $Z_\odot$. As expected, the effects of cooling are highest for the run with initially solar abundance, and lowest in that with pristine gas. We note that the fiducial case (cooling function from Eq.~(\ref{eq:cooleq})) lies between the Sutherland and Dopita cooling curves of pristine and solar for $3\times 10^4 < T < 10^6$~K. The effects of increased cooling are also evident in the amount of gas ejected per unit area. The run with initially pristine gas, for which cooling is least important, has higher mass loading than the run with initially solar abundance gas, for which cooling is most important\footnote{Whilst this paper was in preparation, the EAGLE team \citep{Schaye_2014} also independently suggested a metallicity dependent ejection mass loading for cosmological simulations}. The ejection rates of the intermediate initial metallicity simulation (at $0.1 Z_\odot$) are comparable to those of the fiducial cooling function.

The metal ejection rates of these simulations are a little more complicated. The values of $\beta_{\rm Z}/y$ for the fiducial, pristine, $0.1$ solar and solar simulations are $8.1\%$, $11.9\%$, $13.2\%$ and $34.9\%$ respectively. The solar (highest) metallicity simulation has the highest metal ejection rate, primarily because the ISM gas is already so enriched that the majority of metals ejected are from the ISM, rather than the newly synthesised metals.  After ignoring the effects of the initial gas metallicity (i.e. excluding the average metallicity of the ISM) these figures become $8.1\%$, $11.9\%$, $8.1\%$ and $9.8\%$ respectively. The lower metallicity simulations all have rather similar metal ejection rates, with the fiducial cooling curve having a lower metal ejection rate than the pristine $0.1 Z_\odot$ case. This suggests that there is not a strong trend of metal ejection rate with initial ISM metallicity, i.e. the lowest cooling rate simulation (pristine gas with metal dependent cooling) lies between the fiducial and intermediate metal ejection curves, although we would caution that at this level the ordering may depend on the SN positions (these simulations are a `best case' comparison in that they have the same random seed for SN locations) and averaging over all configurations can adjust the rates. Similarly this ordering also does not suggest a trend with metallicity, except for the high metallicity ISM when the metallicity of the outflow is dominated by the initial ISM metallicity rather than by recently injected metals.

\begin{figure}
\centering
\includegraphics[width=\figwidth]{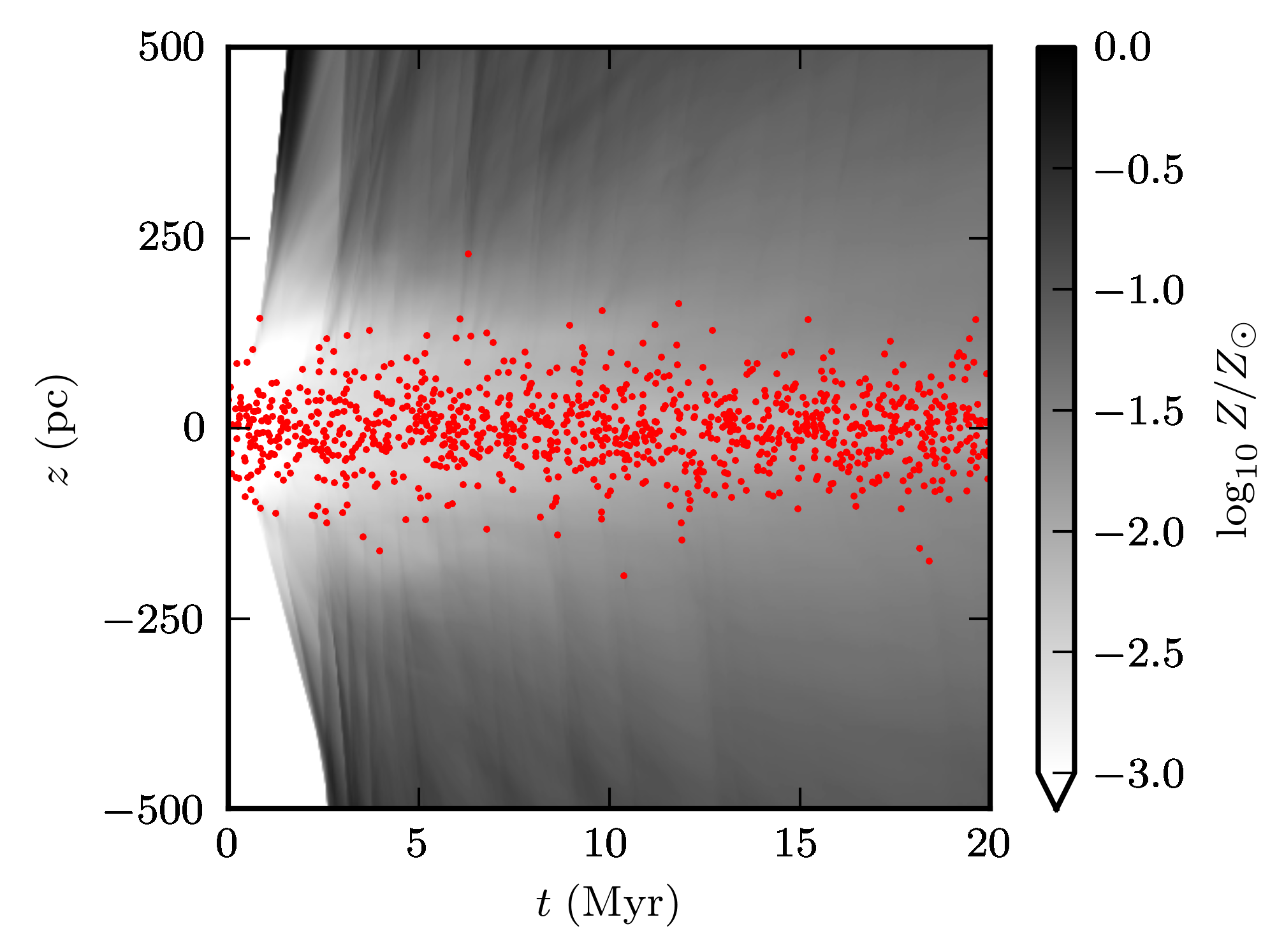}
\caption[Metallicity as a function of height and time]{Mean metallicity as a function of height and time for the simulation in Fig.~1. \emph{Grey-scales} denote the mean metallicity by mass as a fraction of solar, averaged horizontally for each height and time, whilst \emph{red dots} indicate the heights and times of the SNe. During the first 3~Myr the SNe begin to drive a vertical wind from the disk. Individual SNe that are particularly high above the disk can be associated with their individual ejecta, and the discrepancy between the disk and the wind metallicity can be seen.}
\label{fig:time_space}
\end{figure}

In Fig.~\ref{fig:time_space} we show the horizontally averaged mass-weighted metallicity and different heights in the disk for the fiducial parameters. The initially pristine gas gives way to the enriched gas after just a few Myr, and the metals are ejected from the simulation volume. Occasionally (for SNe high above the mid-plane) the individual ejecta can be discerned. After 20~Myr has elapsed the mass-weighted metallicity of the disk is $\sim 10^{-2} Z_\odot$ whilst near the edge of the simulation volume it is around $10^{-1} \, Z_\odot$. There is, however, a correlation between metallicity and velocity and so taking averages at a given height underestimates the wind metallicity, and for unbiased estimates of the wind metallicity in Section~\ref{sec:met_outflow_deps} we use the integrated material ejected.

Varying the properties of the disk does not change the qualitative behaviour of the ISM nor of the outflow. 
As in \cite{Creasey_13} we fit a linear relation to the mass and metal mass outflows as a function of time once the wind has started, to characterise its mass and metal mass loading. We discuss how these quantities depend on disk properties next.

\subsection{Outflow dependencies}\label{sec:met_outflow_deps}
\label{sect:outflow}
\begin{figure}
\centering
\includegraphics[width=\figwidth]{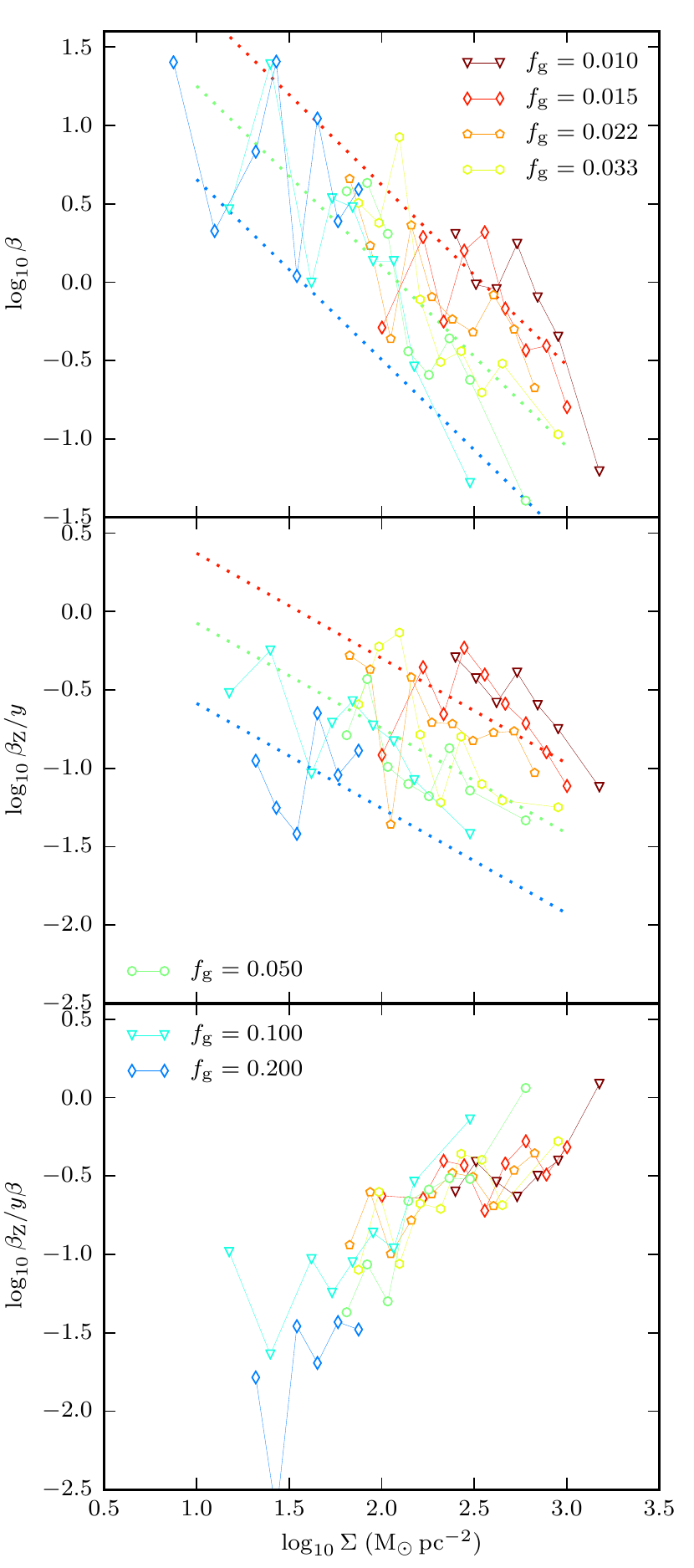}
\caption[Mass and metal mass loading for different surface density simulations.]{The dependence of the mass loading, $\beta \equiv \dot{M}_{\rm w} / \dot{M}_\star$, and the metal mass loading, $\betaZ \equiv \dot{M}_{\rm w,Z} / \dot{M}_\star$, and the ratio $\betaZ/(y\beta)$
({\em top} to {\em bottom}) on the total surface density, $\Sigma = \sigmag / \fg$ (stars and gas). \emph{Coloured symbols} refer to the different gas fractions, $\fg$, as per the legend, power-law fits to the dependence on $\Sigma$ for some values of $\fg$) are overplotted as dotted lines (colours matched to symbol colours). There is a large amount of scatter in the upper two panels due to the stochasticity of the star formation, but definite trends of mass and metal mass loading decreasing with increasing $\Sigma$ and $\fg$ are still clearly visible. The scatter in the bottom panel is less, suggesting that events that induce higher mass ejections also induce higher metal ejections.}
\label{fig:betaZ_sigma_dep}
\end{figure}

In analogy with the mass loading $\beta$ of the wind (Eq.~\ref{eq:beta}), we define the quantity $\betaZ$ as the ratio of the metal mass flux ($\dot{M}_{\rm w, Z}$) over the star formation rate ($\dot{M}_\star$),
\begin{equation}
\betaZ \equiv \frac{\dot{M}_{\rm w,Z}}{\dot{M}_\star} \, ,
\label{eq:betaz}
\end{equation}
as measured from the slopes of the delayed linear fits for the time evolution (i.e. Fig.~\ref{fig:ej_time}) for each simulation. In Fig.~\ref{fig:betaZ_sigma_dep} we explore how $\betaZ$ depends on gas fraction $\fg$, and total surface density $\Sigma = \sigmag/\fg$. Both $\beta$ and $\betaZ$ decrease with increasing gas fraction, and for a given gas fraction, with increasing surface density. This can be understood from the fact that a higher surface density disk exerts a larger gravitational pull on the gas, and it also has a smaller scale-height which increases the cooling rate of the gas. Similarly higher gas fractions also increase the cooling rate, and so in both cases gas that does manage to escape from a dense disk therefore needs to be hotter, and hence such winds have a smaller mass loading. To a large extent the disk surface density can be used as a proxy for galaxy mass, and so we expect {\em higher} mass galaxies to have {\em hotter} winds with {\em lower} mass loading.

A similarly significant trend, somewhat less strong, appears for the metal ejection fraction, i.e. at high disk surface densities and gas fractions the disks are less efficient at ejecting their metals. It should be re-iterated that we are only examining $|z|<500$~pc here, and so these metals may not escape the halo (and indeed may be recycled back in to the disk), an effect which again will be more prominent for more massive galaxies with deeper potentials. 

The ratio $\betaZ/(y\beta)$, plotted in the bottom panel of Fig.~\ref{fig:betaZ_sigma_dep}
is a measure of the average metallicity of the wind. The scatter is reduced in this data, i.e. the SN distributions that are more effective at ejecting metals are also more effective at driving the winds. As discussed in \CTB\ the latter effect seems a result of a greater fraction of SN occurring near to the edge of the disk and is probably the cause of the former too, entailing the correlation. Since the negative trend with surface density is less strong for the metal fraction, the metallicity becomes an increasing function of surface density, i.e. although high density disks are less effective at ejecting metals, they are even less effective at driving a wind and as a result the metallicity of the wind is higher. In summary: denser disks drive hotter, {\em higher} metallicity winds, with {\em lower} mass and metal mass loading.

The best fit regression for the metal ejection fraction in units of the yield $y$ is
\begin{eqnarray}\label{eq:betaZ_bestfit}
{\betaZ\over y} &=& \left[ 0.10 \pm 0.01\right] \left( \frac{\sigmag}{10 \; \rm \Msun \, pc^{-2}} \right)^{-0.67 \pm 0.14}\nonumber\\
&\times & \left( \frac{\fg}{0.1} \right)^{-0.18 \pm 0.10} \, ,
\end{eqnarray}
where we give jackknife errors. We see that the metal mass loading has a negative dependence on gas surface density and a weak dependence on gas fraction. A leading coefficient of $< 0.5$ indicates that the most of the metals distributed by the SNe are retained by the ISM. The best fit for the dependence of the mass loss
is consistent that found in paper I, reported in  Eq.~(\ref{eq:beta_fit}), $\beta\propto \sigmag^{-1.15}\,\fg^{0.16}$. The mass loading has a stronger negative dependence on gas surface density; both depend weakly on gas fraction.  Therefore the metallicity of the wind is $\propto \betaZ/\beta\propto \sigmag^{0.48\pm 0.18}$, implying that winds escaping from higher-surface density disks (or more massive galaxies) have higher metallicity.

It is interesting to contrast these values with limiting cases. If all the ejecta were to escape, but entrain none of the ISM gas, the metallicity of the ejecta would be equal to the yield, $\betaZ =y$ and the mass loading  $\beta = (M_{\rm SN} / 100 \, {\rm \Msun} ) \epsilon_{100}^{-1} = 0.1$. In this case the ISM would remain un-enriched as all the metals are swept out of the simulation volume. We see, however, that in our simulations a large amount of gas is entrained into the wind, $\beta \gg 0.1$, yet most of the ejecta's metals remain in the disk, $\betaZ \ll y$. Nevertheless, the metallicity of the wind still lies between that of the ISM and the ejecta, i.e. $Z_{\rm ISM} < Z_{\rm w} < \Zej$ (see Figures \ref{fig:Z_phase} and \ref{fig:Z_phase_edge}), indeed it would be extremely difficult to alter the order of these metallicities, a point we discuss in the following section.

\subsection{The correlation between enrichment and temperature}
\label{sec:metal_diffuse}
One aspect of the simulations noted in Sec. \ref{sec:ism_metallicity} was the strong correlation between enrichment and temperature. 
The extent to which this is maintained would follow from the balance between the processes that separate these quantities, vs. the processes that cause them to co-locate, or at least have no preference for their separation.

The driving process for co-location in our simulations is that SNe are the sources of all the metals and the major source of additional thermal energy. 
It is also notable that advection will generally transport both metals and thermal energy together.

A slightly more subtle point is that the diffusivity of heat and mass are closely related, due to their underlying molecular origins. 
Whilst diffusion in the most direct sense is not an important process in the ISM, the diffusion that occurs in shocks and also as a result of turbulent mixing very definitely is, and the co-transport of thermal energy and metals in the hydrodynamical solvers is an important effect that causes the metals and thermal energy to co-locate.

In opposition to these are a number of processes which treat the metals and thermal energy separately. Firstly there is radiative cooling, which allows thermal energy but not metals to escape.
In addition, this cooling is in general metal dependent (although only included in a number of our simulations), which generally increases the radiative losses from high metallicity gas. 
Notably cooling is much more effective for the warm, dense gas than it is for the hot, sparse gas, and so we may expect the correlation to be more pronounced in the latter case, and especially in the material that forms galactic winds. 
We should also consider accretion, which will primarily consist of lower metallicity gas whose temperature may be hot (as it shocks against the gas disk), or come in the form of cold streams (see e.g. \citealp{deVoort_2012}).

\begin{figure*}
\centering
\includegraphics[width=2\columnwidth]{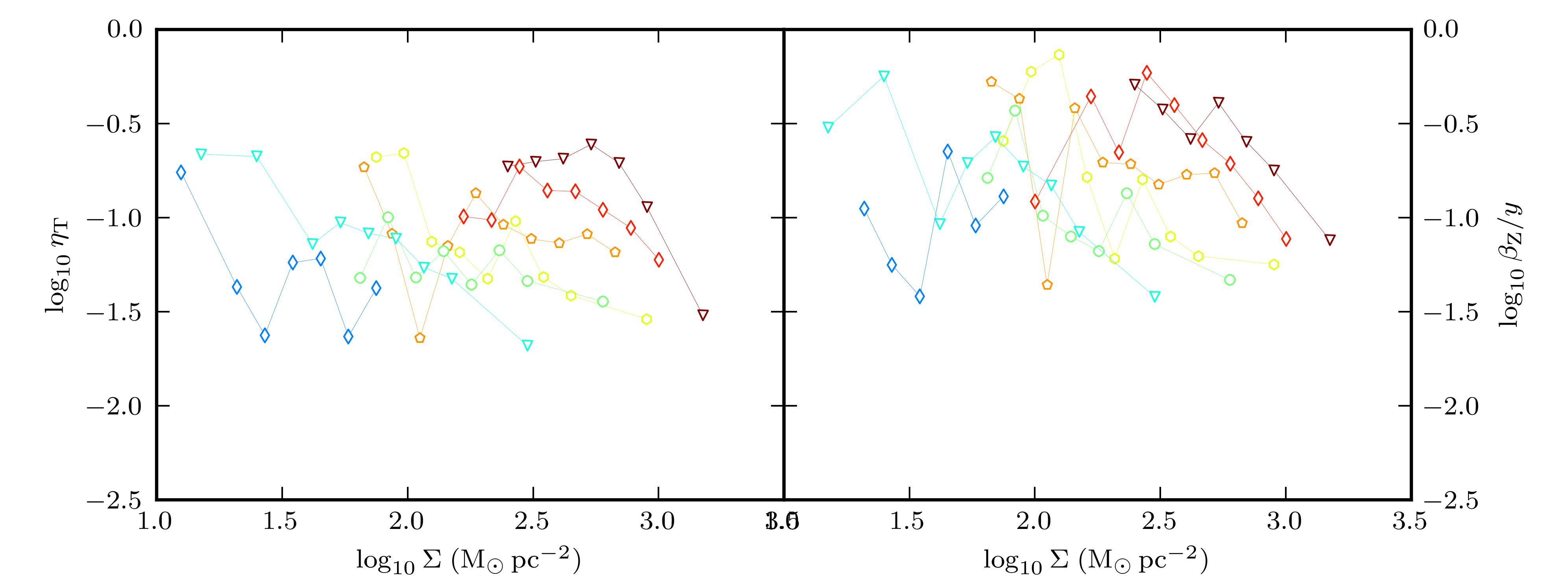}
\caption[The thermalisation efficiency $\eta_T$ (Eq.~\ref{eq:etat}) vs. the fraction of metals entrained $\betaZ/y$ (Eq.~\ref{eq:betaz}) as a function of the total surface density.]{\emph{Left panel}, the thermalisation efficiency $\eta_T$ (Eq.~\ref{eq:etat}) as a function of total surface density $\Sigma$. \emph{Right panel}, the fraction of metals entrained in the wind, $\betaZ/y$ (Eq.~\ref{eq:betaz}). Gas fractions are coloured as for Fig.~\ref{fig:betaZ_sigma_dep}. The dependence of thermalisation efficiency and metal mass loading on the properties of the disk are very similar, with $\eta_T$ smaller than $\betaZ/y$ by $\sim 0.4$~dex due to radiative cooling.}
\label{fig:eta_betaZ}
\end{figure*}

\begin{figure}
\centering
\includegraphics[width=\figwidth]{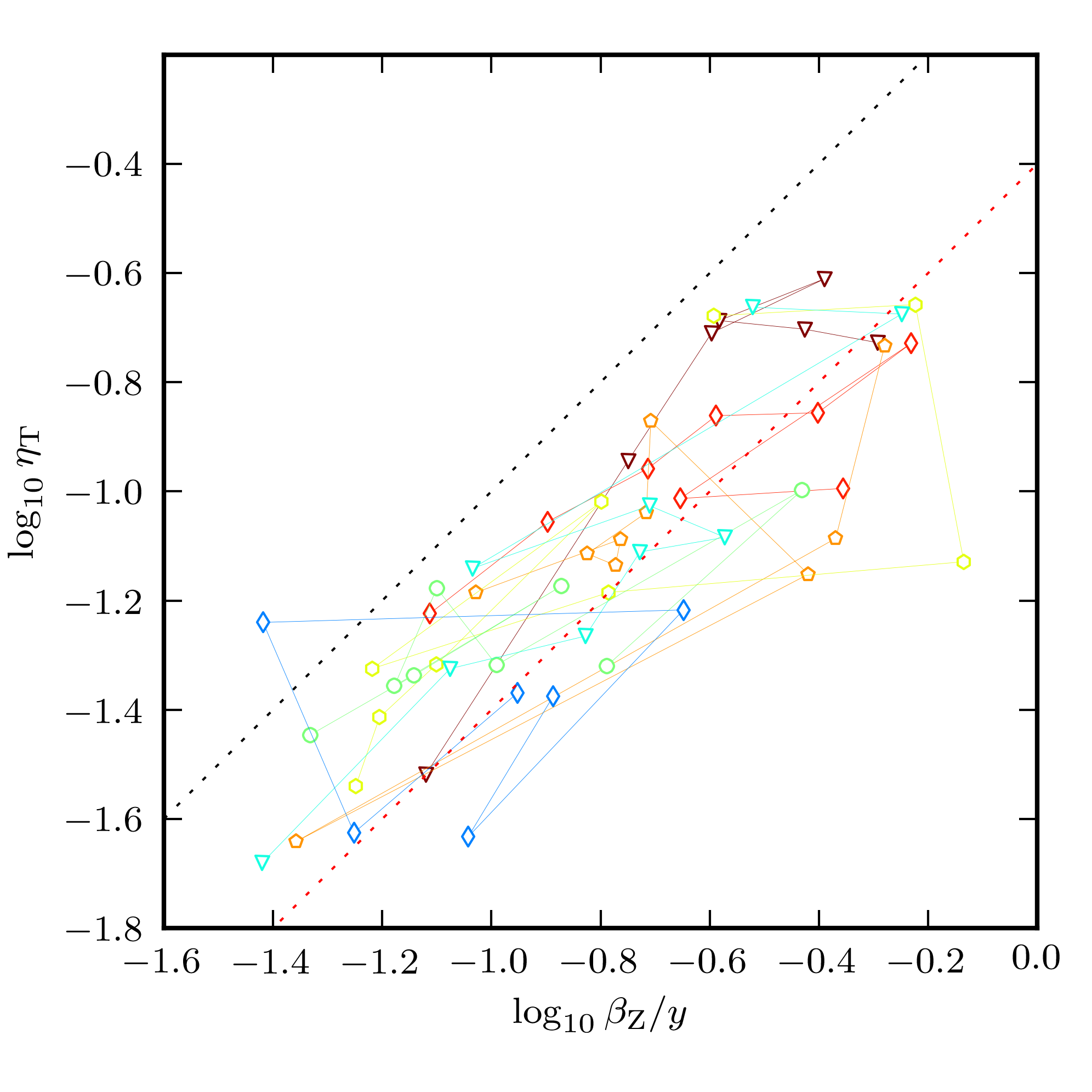}
\caption[The thermalisation efficiency $\eta_T$ vs. the fraction of metals entrained $\betaZ/y$.]{The thermalisation efficiency $\eta_T$ vs. fraction of metals entrained $\betaZ/y$. Gas fractions are coloured as for Fig.~\ref{fig:betaZ_sigma_dep}. \emph{Dotted black line} indicates where the points would lie if $\eta_T=\betaZ/y$, i.e. the same fraction of metals and thermal energy escape. \emph{Dotted red line} indicates the relation $\eta_T=0.4\betaZ/y$, i.e. if the fraction of energy escaping in the outflow is $40\%$ of the fraction of metals ejected.}
\label{fig:eta_betaZ2}
\end{figure}

The preferential distribution of metals in the hot phase also appears to be found in observations, with \citet{Ferrara_2005} finding that only 5\%-9\% of metals lie in the cool ($10^4\; \rm K$)  phase of the ISM, whereas the rest go into an unbound hot phase. Lower limits on the CGM metallicity around Milky-Way sized galaxies \citep{Tumlinson_2011} also suggested they may substantially exceed the fraction of metals in the ISM, consistent with the most recent results from \citet{Werk_2014,Peeples_2014}. Interestingly those latter results include metals at $T=10^4$~K, so whilst they are certainly unbound today, their ionisation state when they exited the ISM is unknown. We explore the correspondence between thermal energy and metallicity in Figures \ref{fig:eta_betaZ} \& \ref{fig:eta_betaZ2}.

In \CTB\ we constructed a model to estimate how efficiently SNe bubbles drive a wind, and what fraction of SNe energy is simply radiated away. The model is based upon the snowplough models of \citet{Cox_72} and \citet{Chevalier_1974}, of a shock expanding in the ISM surrounding the blast, until the thermal energy losses due to cooling become comparable to thermal energy dilution due to shock heating. In this very simple model, a mass $M_{\rm hot}$ of ISM gas was heated by each SN, and this mass escaped to form the galactic wind. Since the blast wave contains all the energy and mass of the SN, we would expect a complete escape of this bubble to also carry $100\%$ of the metals. As we discussed in Section~\ref{sect:outflow}, if only the SNe ejecta themselves were to escape, the resulting mass loading would be $\beta \approx 0.1$. Given that we measure much higher values for the mass loading in these simulations, the amount of gas heated by the SNe must be considerably larger than the 
mass of the ejecta.

Following \CTB, we consider the mean fraction of power `thermalised' in the outflow, $\eta_{\rm T}$ as the sum of the fractions of mechanical and thermal energy in terms of the mean injection rate (by star formation), i.e.
\begin{equation}
\label{eq:etat}
\eta_{\rm T}=\eta_{\rm mech}+\eta_{\rm therm}\,.
\end{equation}
For our simulations where the only other mechanism by which energy can leave the box is radiation (from cooling), then for a steady state disk the sum of $\eta_{\rm T}$ with the mean fraction of energy radiated (e.g. plotted in Fig.~\ref{fig:ej_time}) is unity.

Note that, just as in \CTB, we expect the majority of this energy to be converted into mechanical energy (of the wind) at large distances from the galaxy as the gas adiabatically expands, however inside the ISM the thermal energy is a large component.
Assuming that thermal energy and metals trace each other perfectly, and cooling losses are negligible, then implies $\eta_{\rm T}=\betaZ/y$. Since cooling does play a role, however, we expect $\eta_{\rm T}<\betaZ/y$.  We plot $\eta_{\rm T}$ and $\betaZ/y$ side by side in Fig.~\ref{fig:eta_betaZ}. Their dependence on the properties of the disk are very similar, with some runs with given gas fraction tracking each other in great detail as the surface density is varied, and vice versa, as expected.

In Fig.~\ref{fig:eta_betaZ2} we plot $\eta_{\rm T}$ versus $\betaZ/y$ for those same simulations, and
over plot two scenarios for thermalisation as dotted lines. A model in which the same fraction of metals and energy escapes, which has $\eta_{\rm T}=\betaZ/y$, is shown as the black dotted line. 
A model in which the energy fraction escaping is 40 per cent of the metal fraction that escapes, which has $\eta_{\rm T}=0.4\,\betaZ/y$, is shown as the red dotted line. The latter represents quite a good fit to the simulation results. In a steady-state galactic wind model, one would expect the metals that remain in the disk to radiate all their thermal energy, so this result suggests that the ejecta that escapes radiates away an additional 60 per cent of its energy as it mixes with the ISM gas. Note that the thermalisation $\eta_{\rm T}\le 1$ and $\eta_{\rm T}\le \betaZ/y$: a steady wind cannot carry away energy at a rate that is higher than the SNe energy injection rate, and thermal energy cannot cannot diffuse {\em less} than metals (but it can be lost due to radiation).
It is interesting to compare the thermalisation efficiency to the commonly quoted $10\%$ of \cite{Larson_1974}. Our simulations suggest that this fraction depends upon the metal ejection and is not universal.

\subsection{The effect of stellar winds on mass-loading}\label{sec:wind_effects}
\begin{figure}
\centering
\includegraphics[width=\figwidth]{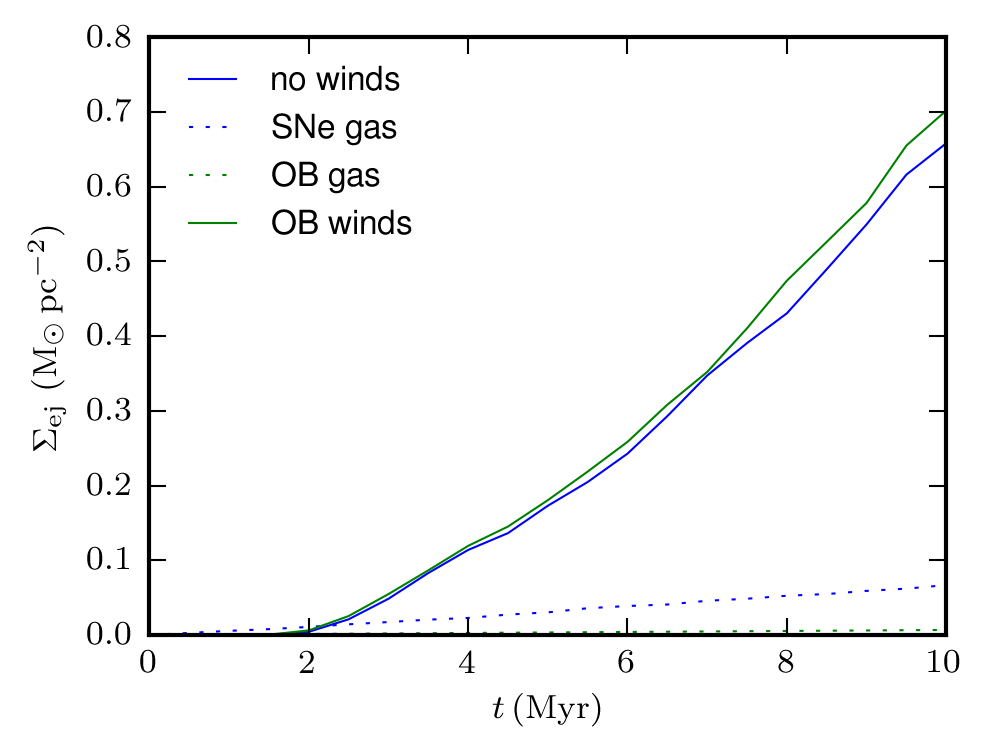}
\caption[Gas ejection due to stellar winds.]{The effect of stellar winds from SNe precursors on a galactic wind
for a model with  star formation rate $\dot{\Sigma}_\star = 6.7 \times 10^{-2} \; \rm \Msun \, kpc^{-2} \, yr^{-1}$.
\emph{Blue solid line} indicates the ejected surface density for a model with thermal energy injection from SNe but neglecting stellar winds,  \emph{green line} is the same model when stellar winds are included as described in the text. The total mass released by SNe and winds per unit area are shown by the dotted lines. The gas ejected from SNe and in winds has coupled to a much larger mass in gas, entraining it in a wind from the disk. The effect of stellar winds is small.}
\label{fig:OB_winds}
\end{figure}

In Fig.~\ref{fig:OB_winds} we study the effects of including stellar winds from OB associations. The overall effect is an increase in the outflow of $\approx 7 \%$, approximately in proportion to their mass as a fraction of the SN ejecta, but far less than their energy input, as the amount of energy we have injected of the 10 million years duration of the stellar winds is of the same order as $E_{\rm SN}$. This may in part be due to our lack of resolution for this process, but our energy injection rate is rather high and the site of the injection (exactly at the position of the SN rather than with a large dispersion) will count against this, so we infer from this that SNe are still the dominant process for gas ejection. 

It is interesting to compare our findings to other simulations in which radiation and dust driving  \citep{Murray_2005, Martin_2005, Sharma_2011} were included, such as those described by \cite{Hopkins_2011a} and applied in \cite{Hopkins_2012}. The latter paper concludes that although radiation and winds are important for unbinding giant molecular clouds, SNe are still the dominant mechanism for driving galactic winds. This is consistent with the results in Fig.~\ref{fig:OB_winds}. Our simulations are less reliant on the stellar winds to destroy the dense star forming regions since our SN distribution is pre-computed according to the Kennicutt-Schmidt relation and not tied to the gas distribution in which they explode. If anything we see even less effect of stellar winds on the outflow, which may be due to Hopkins' use of multiple photon scatterings to further increase the momentum imparted by the radiation. 
We note that this system is relatively gas-poor compared to the high redshift galaxy progenitors that \citet{Hopkins_2012} suggested may be affected most, however we did not try these cases as we were not convinced that our simulations could correctly follow the evolution of the HII regions at higher gas surface densities, where the higher recombination rates can suppress the formation of HII regions at (even our) grid resolution.  We leave a more thorough analysis of the parameter space of wind driving on these scales for a future paper. Our placement of SNe in environments that are not necessarily the densest may also cause us to underestimate the importance of winds, nevertheless in our simulations it is undoubtedly realistic since massive stars do not explode in dense environments.

Fig.~\ref{fig:OB_winds} also provides an alternative aspect from which to imply the mass loading of the winds. Since the gas ejection rate from SNe is related to the star formation rate via the ejecta mass and the proportion of SNe in the IMF, i.e. $\dot{\Sigma}_{\rm SN} / \dot{\Sigma}_\star = (M_{\rm SN} / 100 \, {\rm \Msun} ) \epsilon_{100}^{-1} \approx 0.1$ for this model, then the mass loading can be deduced as the product of this and the ratio of ejected surface density to SN ejecta. This can be read from Fig.~\ref{fig:OB_winds} as $\dot{\Sigma}_{\rm w} / \dot{\Sigma}_{\rm SN} \approx 10$, implying $\beta \equiv \dot{\Sigma}_{\rm w} / \dot{\Sigma}_\star \approx 1$ for this simulation.

\subsection{Summary}
In this section we have analysed a series of numerical simulations of different galaxy disk environments with and without stellar winds in order to understand the effects on the enrichment of the outflows (the effect on the mass loading being the focus of \CTB).
The primary result of these simulations is that higher surface density disk environments display more enriched outflows, and indeed the metallicity of the outflow is higher than the metallicity of the disk. 
This is compatible with the the outflows having lower mass (i.e. a lower `mass loading') since SNe are the source of both metals and thermal energy, however this is not generally considered in models of galaxy formation. 
As such, in Section \ref{sec:mass_metallicity} we construct a simple model of galaxy growth that incorporates outflows whose metallicity depend on galaxy mass.

\section{Application: The Mass Metallicity relation}\label{sec:mass_metallicity}
In this section our aim is to use the simulation results to understand the evolution of the mean gas phase metallicity $\left< \Zg \right>$ of a galaxy, and the mass-metallicity relation of galaxies ($\Zg(M_\star)$). We construct a simple model of metallicity distributions from the gas and stellar evolution of galaxies along with the metal ejection rates. We demonstrate that these assumptions are consistent with the metallicity distribution of faint stars (no G-dwarf \lq problem\rq) and that it is straightforward to match the observed mass-metallicity relation of galaxies with reasonable values for the metal ejection rates. We discuss the origin of the turnover in the mass metallicity relation in our model and compare to other analytic models. Finally, we compare the metallicity ejection rates with those found in the hydrodynamical simulations of Section \ref{sec:betaZResults}.

\subsection{Inferring ISM metallicities}\label{sec:inf_metal}
The gas reservoir, $\Mgas$, of a galaxy evolves with the following sources and sinks:
\begin{equation}\label{eq:Mg_evol}
\dot{M}_{\rm g} = \dot{M}_{\rm a} - (1-f_r ) \dot{M}_\star - \dot{M}_{\rm w} \, ,
\end{equation}
where $\dot{M}_{\rm a}$ is the (cold) gas accretion rate, $\dot{M}_\star$ is the star formation rate, $ \dot{M}_{\rm w}$ is the wind loss rate and $f_r$ is the fraction of gas released back in to the ISM via short lived stars and stellar winds (we assume instantaneous recycling\footnote{Note that in our convention $M_\star$ refers to the total amount of stellar mass created, not just the fraction $(1-f_r) M_\star$ that is in long-lived stars.}). The total metal mass of this gas reservoir, $M_{\rm Z}$, evolves as
\begin{eqnarray}
\dot{M}_{\rm Z} &\equiv & \frac{\rm d}{{\rm d}t} \left[ \Zg \Mgas\right] \\
&=& (y+f_r \Zg) \dot{M}_\star + Z_{\rm a} \dot{M}_{\rm a}  - Z_{\rm w} \dot{M}_{\rm w} - \Zg \dot{M}_\star \label{eq:Mz_evol}\, ,
\end{eqnarray}
where $y$ is the yield and $Z_{\rm a}$, $Z_{\rm w}$ and $\Zg$ are the metallicities of the accreting, wind and ISM gas respectively. The terms on the right then refer to metals released by short lived stars, metals accreted from inflowing gas, metals lost in the wind and metals locked away by star formation, respectively.

Our first approximation is to set $Z_{\rm a}=0$, assuming that the metallicity of the inflowing gas is negligible compared to that of the ISM. We note that this may be violated for high mass galaxies that are recycling metals through their halos and we return to this in Section \ref{sec:sim_compare}. We also write the outflowing metals in terms of the star formation rate, $ Z_{\rm w} \dot{M}_{\rm w} \equiv \betaZ(t, M_\star) \dot{M}_\star$, transforming Eq. (\ref{eq:Mz_evol}) to
\begin{equation}\label{eq:Zg_evol}
\dot{Z}_{\rm g} \Mgas + \Zg \dot{M}_{\rm g} = \left[ y-\betaZ (M_\star, t)  - \Zg (1-f_r) \right] \dot{M}_\star \, .
\end{equation}
If we then make the assumption that there is no extra dependence on time (or equivalently redshift), other than that implied in the stellar mass, i.e.
\begin{eqnarray}
\Mgas (M_\star, t) &=& \Mgas (M_\star) \\ 
\Zg (M_\star, t) &=& \Zg (M_\star) \\
\betaZ (M_\star, t) &=& \betaZ (M_\star) 
\end{eqnarray}
then we can write Eq. (\ref{eq:Zg_evol}) parameterised in terms of $M_\star$ rather than time as
\begin{equation}\label{eq:Zg_evol_mstar}
 \frac{\rm d}{{\rm d} M_\star} \left( \Zg \Mgas\right) = -(1-f_r)  \Zg + y - \betaZ (M_\star) \, .
\end{equation}

In Appendix \ref{sec:closedbox} we use this relation to calculate the metallicity distribution function of stars, and show that is does not suffer from the large tail to very low metallicities of the (simplistic) \lq closed box\rq\ model. 

In order to proceed further we need some estimate of $\Mgas (M_\star)$, i.e. the gas mass the corresponds to a given stellar mass. In effect we are assuming that galaxies move along the redshift zero relations, in which case the evolution is parameterised by stellar mass. In general this precludes growth in stellar (and gas) mass due to mergers, which would disrupt this relation, however for lower-mass galaxies mergers are not significant. This assumption results in a mass-metallicity relation that is independent of redshift and there is some evidence that this is indeed the case if one takes a homogeneous sample (see e.g. \citealp{Stott_2013}).

We will assume that gas and stellar mass are related as
\begin{equation}\label{eq:mgass_mstar}
{\Mgas\over 10^{10}\Msun} = A_{\rm g} \left( \frac{M_\star}{10^{10} \; \rm \Msun} \right)^{\alpha_{\rm g}} \, ,
\end{equation}
(we discuss the observational $\Mgas$~-~$M_\star$ relation later in this section), where $A_{\rm g}$ and $\alpha_{\rm g}$ are dimensionless constants. This allows us to solve Eq. (\ref{eq:Zg_evol_mstar}) in integral form as
\begin{eqnarray}\label{eq:Zg_integ}
\Zg\Mgas&=&  \int_0^{M_\star} {\rm d}m \left( y - \betaZ(m) \right) \nonumber \\
&\times& \exp \left[ - \left( \frac{1-f_r}{1-\alpha_{\rm g}} \right) \left( \frac{M_\star}{\Mgas(M_\star)} - \frac{m}{\Mgas(m)}\right) \right] ,
\end{eqnarray}
where $M_\star$ is now the independent variable.
\subsubsection{Special cases of $M_\star \gg \Mgas$ and  $M_\star \ll \Mgas$}
Equation (\ref{eq:Zg_integ}) has two special cases that are of particular interest when approximating the mass-metallicity relation, that of $M_\star \gg \Mgas$ and  $M_\star \ll \Mgas$, corresponding to high and low stellar mass galaxies. In the low gas reservoir limit, we find
\begin{eqnarray}
\Zg &=& \lim_{\Mgas/M_\star \to 0} \frac{1}{\Mgas(M_\star)} \int_0^{M_\star} {\rm d}m \left( y - \betaZ(m) \right) \nonumber \\
&& \exp \left[ - \left( \frac{1-f_r}{1-\alpha_{\rm g}} \right) \left( \frac{M_\star}{\Mgas(M_\star)} - \frac{m}{\Mgas(m)}\right) \right] \nonumber \\
&=& \frac{y - \betaZ(M_\star)}{1-f_r} \, , \label{eq:Zg_equil}
\end{eqnarray} 
i.e. instantaneous response of the metallicity to the star formation. The gas reservoir is so small that it no longer has no `memory' of the star formation history, and the metallicity is set purely by the yields and the current metal ejection rate. Eq. (\ref{eq:Zg_equil}) is also the limit when star formation is \emph{slow}, i.e. Eq. (\ref{eq:Zg_evol}) is allowed to evolve to a point where $\dot{Z}_{\rm g}=\dot{M}_{\rm g} = 0$ (which will be violated for small galaxies since the total stellar mass is increasing fast).

In the gas dominated case, $M_\star \ll \Mgas$, Eq.~(\ref{eq:Zg_integ}) yields
\begin{eqnarray}
\Zg(M_\star) &\approx & \frac{M_\star}{\Mgas} \left( y - \betaZ(M_\star) \right) \, , \label{eq:mstar_small}
\end{eqnarray}
to first order in $y - \betaZ(M_\star)$. In this limit the mass-metallicity relation is determined by the evolution of gas and stellar mass. The metallicity is not set by an equilibrium, but rather is set by the cumulative yield of the total mass in stars that have been formed up to that time.

It is notable that these relations do not depend \emph{directly} either upon the gas accretion rate $\dot{M}_{\rm a}$ nor on the mass loading of the wind, $\beta = \dot{M}_{\rm w} / \dot{M}_\star$, i.e. the metallicities derived from Eq. (\ref{eq:Zg_integ}) do not include those terms. They are of course included implicitly, however, as they are the galactic evolution processes that shape the evolution of the gas mass and thus the $M_\star$-$\Mgas$ relation in Eq. (\ref{eq:mgass_mstar}). 

One final case that is of interest is when we assume ISM gas and metal masses have reached an equilibrium, i.e. Eq. (\ref{eq:Mz_evol}) with the LHS set to zero, and that the metallicity of the outflow $Z_{\rm w} = \Zg$, the mean ISM metallicity. This corresponds to a well mixed outflow, with equilibrium metallicity of outflow and ISM equal to 
\begin{equation}\label{eq:Dave_met}
\Zg = \frac{y}{1+\beta - f_r} \, .
\end{equation}
The case with $f_r=0$ is discussed in FD08 and \citet{Dave_2011}.

\subsubsection{Relating stellar and ISM gas masses}
\label{sect:gas_to_star}
To solve Eq.~(\ref{eq:Zg_integ}) in the general case requires knowledge of the stellar mass to ISM gas mass relation. Observationally, stellar masses are usually inferred from the K-band luminosity (e.g. \citealp{Bell_2001}); however, inferring the gas mass is more subtle. \cite{McGaugh_2005} and \cite{West_2009} give stellar mass to HI (21cm) gas masses, but for high mass galaxies there may also be a large $\rm H_2$ component. \cite{Leroy_2008} estimate the HI+$\rm H_2$ mass using CO measurements. For our purposes some of the HI may not be important for the ISM, as the HI disk is considerably more extended than the stellar disk \citep{Walter_2008}. These data sets are also primarily focused on star forming galaxies, which bias their normalisations to higher gas masses \citep{Catinella_2010}, and the gas fractions could be a factor of 2 smaller (see e.g. \citealp{deRossi_2013}).

A more indirect method of measuring the total gas mass is via the star formation rate (inferred from the H-$\rm \alpha$ luminosity), e.g. as used by \cite{Tremonti_2004}. Some radial profile for the gas surface density is assumed and then a normalisation deduced by inverting the Kennicutt-Schmidt relation.

The use of the above methods to construct a stellar mass to gas mass relation is discussed in some detail in \cite{Peeples_2011}, where the gas mass seems to be well fit by a power law in stellar mass. Gas mass increases with stellar mass, but in a less than proportionate way, so the gas fraction is a decreasing function of stellar mass. Ignoring the considerable scatter, a good fit the HI and $\rm H_2$ data
is of the form given by Eq.~(\ref{eq:mgass_mstar}) with $A_{\rm g}=1$ and $\alpha_{\rm g}=1/2$ \citep[e.g.][]{Peeples_2011,Pap_2012}. The gas masses found from inverting the Kennicutt-Schmidt relation would prefer a higher exponent, nearer to $\alpha_{\rm g}\sim 0.8$, i.e. the gas masses found from integrating HI over the galaxy disks exhibit a much weaker dependency upon stellar mass than the gas mass inferred from the star-formation rate. These quantities can be discrepant because the the star formation is preferentially tracing the molecular gas \citep[e.g.][]{Leroy_2008} and indeed the HI sizes are much more extended than the star forming disk \citep{Walter_2008}, and thus we choose the HI estimates as a better tracer of the ISM gas which will dilute the metals.

\subsection{Predicting the Mass-Metallicity relation}
In this section we apply the results from our simulations in addition to the analysis of section \ref{sec:inf_metal} in order to deduce the mass metallicity relation of galaxies. This allows direct comparison between simulated metallicities and those from observations.

In this section we will compare our results to the  \cite{Kewley_2008} fits to the $M_\star$-$\Zg$ data of \cite{Tremonti_2004} and \cite{Denicolo_2002}, 
\begin{eqnarray}
12 + \log_{10} \left( {\rm O}/{\rm H}\right)_{\rm T04} = 8.566 + 0.475 x - 0.095 x^2 -0.003 x^3 \, , \label{eq:T04} \\
12 + \log_{10} \left( {\rm O}/{\rm H}\right)_{\rm D02} = 8.491 + 0.349 x -0.102 x^2 + 0.008 x^3 , \label{eq:D02}
\end{eqnarray}
(see also \citealp{Peeples_2011}) where $x \equiv \log_{10} \frac{\rm M_\star}{10^9 \; \Msun} $ and T04 and D02 refer to \cite{Tremonti_2004} and \cite{Denicolo_2002}, respectively\footnote{Note our definition of $x$ differs from that of \cite{Kewley_2008, Peeples_2011}, to expand about a galactic stellar mass of $10^9 \; \rm \Msun$, which alters the coefficients and truncates the rather large number of digits required.}. In general there is a scatter in the metallicity data of a factor $\sim 2.5$ in the mass range $10^9 - 10^{11} \; \rm \Msun$ \citep{Kewley_2008}, where the T04 relation has a relatively steep slope compared to other fits. D02 has a mid-range slope but is slightly low in normalisation, by a factor $\sim 1.5$, in comparison to the distribution of fits displayed in \cite{Kewley_2008}.

These are converted to metal mass fractions using
\begin{equation}
\log_{10}  \Zg =\log_{10} {\rm O}/{\rm H}+0.9560 \, ,
\end{equation}
i.e. the Oxygen is assumed to be polluting a primordial mix that was approximately $75\%$ H by mass.

We now turn our attention to the $M_\star$-$\Zg$ relation predicted by our analytic model. By combining the metallicity integral in Eq. (\ref{eq:Zg_integ}) with the gas mass to stellar mass relation from Eq. (\ref{eq:mgass_mstar}) with $A_{\rm g}=1$, $\alpha_{\rm g}=1/2$, the only remaining component is the dependency of the metals lost in the wind, $\betaZ$, upon the stellar mass. This quantity was calculated for small patches of a disk in Section \ref{sec:met_outflow_deps}, and we will explore this correspondence in Section \ref{sec:sim_compare}. We begin, however, by exploring some simple models for the dependence on stellar mass that illustrate how we can compare to the observed $M_\star$-$\Zg$ relations.

We now construct some fits for the retained metal fraction. Recall that $\betaZ$ is defined as the mass of metals ejected from the disk (not necessarily the halo) per unit mass of star formation, so $1-\betaZ/y$ is the fraction of metals retained, hence lies in the range $[0,1]$. We now construct a series of fits where the retained metal fraction is approximated by a power law, 
\begin{equation}\label{eq:ejec_pl}
1- \frac{\betaZ}{y} = \left\{ \begin{array}{cc} 
 \FR \times 10^{-\fR x} \, ,&  M_{\star} \geq \FR^\frac{1}{\fR} \times 10^9 \; {\rm M}_\odot  \\
1 \, , & M_{\star} \leq \FR^\frac{1}{\fR} \times 10^9 \; {\rm M}_\odot  \,,  \end{array} \right.
\end{equation}
i.e. the stellar mass to the power $-\fR$, where we have assumed that $\FR \in [0,1]$ and $\fR > 0$. Since a two-parameter fit allows a considerable amount of freedom for our data sets, we also consider the single parameter (constant fit) of only $\FR$ (with no limitations on the domain).

\begin{table}

\begin{center}
\begin{tabular}{ lccc}
& & $\FR$ & $\fR$ \\
\hline
Rcon &\vline & $0.62$ & $0$ \\
R1 &\vline & $0.64$ & $0.02$ \\
R2 &\vline &  $0.47$ & $0.10$ \\
R3 &\vline & $0.42$ & $0.20$ \\
\end{tabular}

\caption{Parameters to the power-law fits, described in Eqn. (\ref{eq:ejec_pl}). Rcon is the single parameter constant fit (i.e. $\fR=0$) to the T04 relation, whereas R1, R2 and R3 are the two parameter fits to the T04, D02 and HiZELS metallicity relations, respectively.}
\label{tab:r_fits}
\end{center}
\end{table}

In Table \ref{tab:r_fits} we have fitted the power law for the retained metal fraction to the fits for the Eqns. (\ref{eq:T04})  and (\ref{eq:D02}), along with a fit to the  $z=0.84$-$1.47$ HiZELS data points from \cite{Stott_2013}, which we refer to as R1-3. Additionally we include the single-parameter fit to Eqn. (\ref{eq:T04}) which we refer to as Rcon. Each fit is the least squares error fit to the four stellar mass values of the HiZELS data, which gives an approximately uniform coverage of the stellar mass range.

\begin{figure}
\centering
\includegraphics[width=\figwidth]{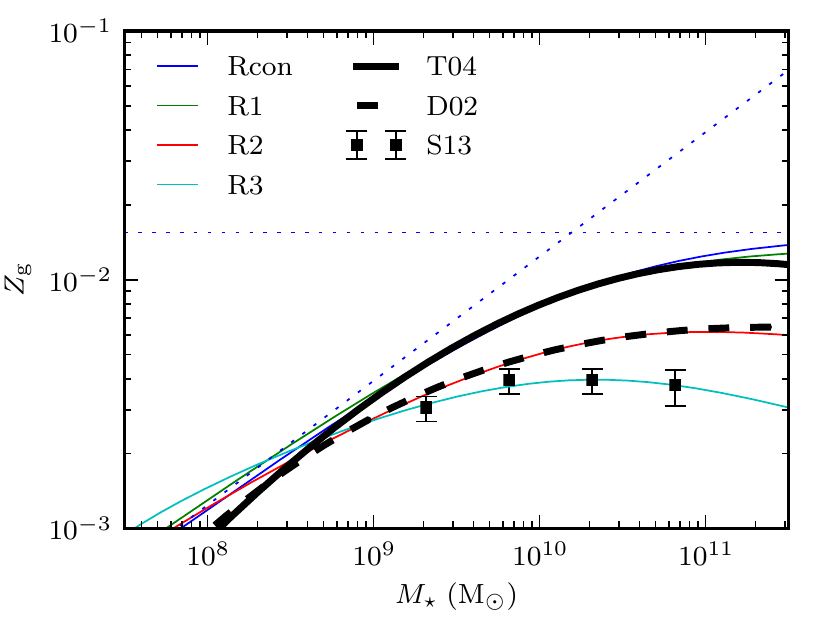}
\caption[Derived mass-metallicity relations for the different models of metal outflows.]{Derived mass-metallicity relations for the different models of metal outflows. \emph{Heavy black solid} and \emph{dashed} lines are the \cite{Tremonti_2004} and \cite{Denicolo_2002} $M_\star$-$\Zg$ relations, respectively. \emph{Points with error bars} refer to the $z=084-1.47$ data of \cite{Stott_2013}. \emph{Blue, green, red} and \emph{cyan solid} lines correspond to the our model using the parameters for the retained metal fractions given in Table \ref{tab:r_fits} respectively. \emph{Blue dotted lines} are the asymptotic relations for low and high gas fractions for the Rcon fit (\emph{blue solid line}) in Eqns. (\ref{eq:asym_low}) and (\ref{eq:asym_high}).
}
\label{fig:pred_metals}
\end{figure}

In Fig.~\ref{fig:pred_metals} we show the effects of these retained fractions against the observed mass metallicity relations. We can immediately see that we have achieved good normalisations without invoking extreme retained fractions. The normalisation is degenerate between changing the yield, $y$, and normalisation of the retained fraction, but a value of $\approx 50 \%$ for the fraction of metals ejected is quite reasonable (and consistent with, say, the IGM calculations of \citealp{Ferrara_2000} or the QSO absorption studies in \citealp{Werk_2012}).

In addition to the normalisation, the transition from large to small (or even negative) slope also seems to fit well. As the exponent of $M_\star$ in the retained fraction falls, e.g. to the most in extreme value R3, the slope of the $M_\star$-$\Zg$ falls correspondingly at high masses. This can be understood in terms of the special cases described in Section \ref{sec:inf_metal}, and are shown as dotted blue lines in Fig.~\ref{fig:pred_metals}.

At low stellar masses, we will be in a gas dominated phase, $M_\star \ll \Mgas$, and hence from Eq. (\ref{eq:Zg_integ})
\begin{eqnarray}
\Zg & \approx & \frac{M_\star}{\Mgas} \left( y - \betaZ(M_\star) \right) \label{eq:asym_low} \\
 & \approx & \left( \frac{M_\star}{10^{10} \, \rm \Msun}\right)^{1/2} \left( y - \betaZ(M_\star) \right) \, ,
\end{eqnarray}
The slope of the $\Zg-M_\star$ relation is then set almost entirely by the gas mass to stellar mass relation, which for the parameterisation of Eq.~(\ref{eq:mgass_mstar}) is
$\Zg \propto (M_\star/M_{\rm gas}) \approx M_\star^{0.5}$ for the value of $\alpha_g=0.5$. This is the sloping blue dotted line in Fig.~\ref{fig:pred_metals}. As \citet{Peeples_2011} argues, these lower values of $\alpha_g$ (as opposed to values closer to 1 found by inverting the K-S relation) are more appropriate here since the metals are diluted by the total gas rather than the molecular gas traced by the star formation. Notably \citet[after this paper was submitted]{Zahid_2014} similarly argues that the turnover mass is driven by saturation of the stellar fraction (of baryonic material) for galaxies at $z \leq 1.6$.

At the high stellar masses the gas reservoir is small, $\Mgas \ll M_\star$, and we have the limit given by Eq. (\ref{eq:Zg_equil}),
\begin{equation}
\Zg \approx \frac{y - \betaZ(M_\star)}{1-f_r} \label{eq:asym_high} \, .
\end{equation}
The metallicity simply follows the retained fraction slope which is more gentle: it is the horizontal blue dotted line in Fig.~\ref{fig:pred_metals}. We note that this gentle slope requires that the retained fraction must be a weak function of stellar mass, with exponents of $-0.1$ and $0$ for the relations shown.

The transition of high to low gas fraction with increasing $M_\star$ in our model causes the turnover in the slope of the mass-metallicity relations in Fig.~\ref{fig:pred_metals}, which gradually transitions as the stellar mass exceeds the gas mass. 
Whilst not being inconsistent with the standard explanation of the turnover (that it is due to galactic winds, see e.g. \citealp{Tremonti_2004}), this description does provide additional nuances that make it distinct from other models.

One such alternative is given in FD08, where it is argued that the transition in slope is driven by the changes in mass loading from high ($\beta \gg 1$) to low ($\beta \ll 1$) with increasing $M_\star$, and that the metallicity is approximately given by Eq.~(\ref{eq:Dave_met}). 
In the FD08 model it is assumed that the galaxies are in instantaneous equilibrium, with inflow balancing star formation and outflow, which allows the deduction of the dilution of metals and hence the gas phase metallicity (in equilibrium). In our model we assume a relation between the gas mass and the stellar mass and hence at all stellar masses we can calculate the dilution. Since our galaxies are growing in gas mass whilst the FD08 gas reservoir is constant, the metallicities will be distinct, although for the high mass galaxies the growth rates will be small and hence our predictions similar. Even for low mass galaxies, however, if the outflow rate ($\dot{M}_{\rm w}$) is an approximately constant fraction of the inflow rate ($\dot{M}_{\rm a}$), then for power-law inflow rates (in $M_\star$) the ratio of the gas mass to the stellar mass compared to the mass loading will only differ by a factor, making the models very similar in form and normalisation. This explains why both models give similar $\Zg$~-~$M_\star$ relations even though the underlying assumptions are quite distinct.

In terms of retained fraction of metals, the FD08 model assumes a well-mixed ISM, i.e. the wind metallicity is the same as the ISM metallicity, in contrast to our model where the wind can preferentially carry away metals. This allows our models slightly more freedom to adjust the slope of of the $M_\star$-$\Zg$ relation at high stellar masses, where the mass loading $\beta \ll 1$, i.e. Eq. (\ref{eq:Dave_met}) implies the metallicities will converge to the effective yield, whereas the introduction of $\betaZ(M_\star)$ in  Eq. (\ref{eq:Zg_equil}) releases us from this constraint.

\subsection{Comparison with simulations}\label{sec:sim_compare}
\begin{figure}
\centering
\includegraphics[width=\figwidth]{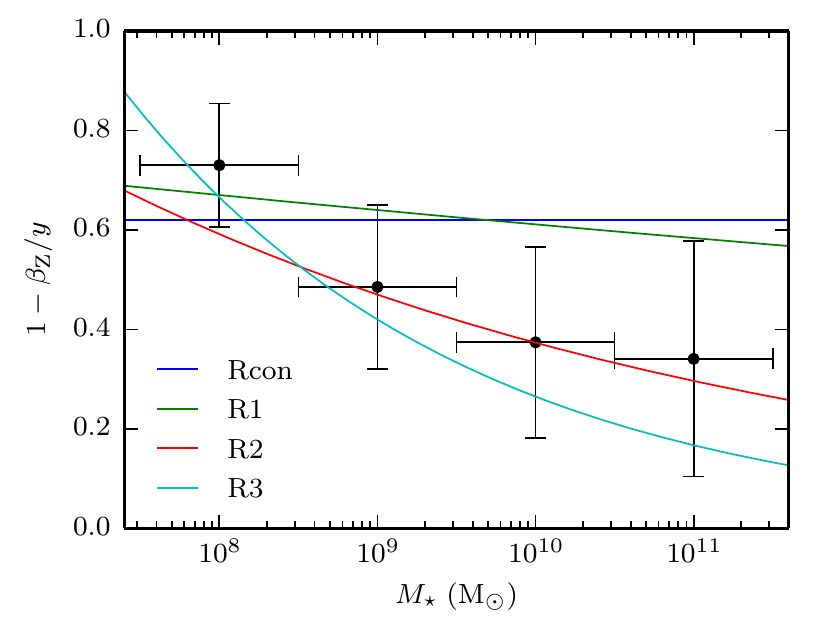}
\caption{
Estimates for the retained metal fraction, $1 - \betaZ / y$ as a function of stellar mass. 
\emph{Blue, green, red} and \emph{cyan} lines are the power-law models from Table \ref{tab:r_fits} (and Fig.~\ref{fig:pred_metals}) whose parameters were fitted to match the observed data. 
\emph{Black points with error bars} denote bins of the simulation outputs (see Section \ref{sec:betaZResults})  binned into stellar masses of $10^8$, $10^9$, $10^{10}$ and $10^{11} \; \rm \Msun$, see main text for details of the conversion.
}
\label{fig:ret_metal}
\end{figure}

We are now in a position to discuss perhaps the most interesting aspect of all, the use of hydrodynamical simulations which in principle allow us to predict the retained metal fraction and compare it to observations. We should bear in mind, however, that the fits to the observed mass-metallicity relation all have a rather weak dependency of retained metal fraction as a function of stellar mass, i.e. the exponents $-\fR$ lie in the range $[-0.2, 0]$.

The simulations performed in Section \ref{sec:betaZResults} were parameterised in terms of gas surface density $\sigmag$ and gas fraction $\fg$, so to put these on the mass metallicity relation we must transform these into stellar masses. To do so we perform a similar analysis as in \CTB\ to estimate dependencies on circular velocity from those on surface density. We begin by assuming the disks to be exponential with total mass 
\begin{equation}\label{eq:md_sigmag}
M_{\rm d} = 2 \pi \Sigma R_{\rm d}^2 \, ,
\end{equation}
where $R_{\rm d}$ is the disk scale length, and $\Sigma_{\rm g}$ the central gas surface density. \cite{Shen_2003} estimates the sizes of disks to scale weakly with stellar mass as $R_{\rm d} \propto M_\star^{0.15}$.  We normalise to the Milky Way, assumed to have a stellar mass of $5\times 10^{10} \; \rm \Msun$ and scale radius $2.5 \; \rm kpc$. Putting the total mass of the disk in Eq.~(\ref{eq:md_sigmag}) together with the power-law galaxy stellar-mass to size relation gives the conversion
\begin{eqnarray}\label{eq:mstar_sigmag}
M_\star &=& \frac{1-\fg}{0.9} \times \left( \frac{\Sigma_{\rm g} \fg^{-1} }{1270 \, \Msun \, {\rm pc}^{-2}} \right)^{1+2\times 0.15} \times \nonumber \\
&&   \times 5 \times 10^{10} \, \Msun \, ,
\end{eqnarray}
i.e. galaxies with low surface densities (or just high gas fractions $\fg$) correspond to low stellar mass galaxies, whilst at higher surface densities or very low gas fractions the galaxies have higher stellar masses.

Combining all our simulations - parameterised by $\sigmag$, and integrating these over exponential disk, we are in a position to compute the retained metal fraction, $1-\betaZ/g$, as function of stellar mass, shown in Fig.~\ref{fig:ret_metal}, where we have combined the data into stellar mass bins of $10^{7.5 {\rm -} 8.5}$, $10^{8.5 {\rm -}9.5}$, $10^{9.5 {\rm -} 10.5 }$ and $10^{10.5 {\rm -} 11.5} \; \rm \Msun$. 
For each bin we plot the mean retained fraction and the error bars, which due both to the stochastic nature of the energy injection by SNe (in time and in position in the disk), and partly due to the different gas fractions and surface densities that correspond to a single stellar mass.

Noticeably the weak trend in simulation data, i.e. lower retained metal fractions (high metal ejection fractions) with increasing stellar mass (i.e. at higher total surface density) appears in tension with the fit in Eqn.~(\ref{eq:betaZ_bestfit}), where metal ejection fractions fell with higher gas surface densities (but fell with gas fractions). This inversion is partly due to the conversion from logarithmic to linear space, and partly due to the mapping of the parameters domain on to stellar mass. 
The contribution of the former is due to the increased variance between the simulations as a function of stellar mass, i.e. it causes the mean to rise when we take the (convex) map from logarithmic to linear.
The mapping of the parameters matters because there is a lack of high gas fraction high stellar mass galaxies, and low gas fraction low stellar mass galaxies. Whilst this limitation is qualitatively realistic - those objects are indeed more rare than the parameters we explore - this does highlight the importance of the galaxy distribution and that the trends for the simulation points in Fig.~\ref{fig:ret_metal} should really only be considered in a qualitative sense.

Along with our simulation data we have plotted the power-law models in Table~\ref{tab:r_fits} of the retained metal fraction.
Although we argue it is unwarranted to place a great deal of attention to the exact normalisation of our relations - due to both the outliers, uncertainties in yields, and the $\Zg$-$M_\star$ relation etc.- the metal retention factors that results from the combination of simulations and the assumption of an exponential disk do appear reasonable, both between the curves and the data and compared to IGM estimates (e.g. \citealp{Ferrara_2000, Werk_2012}). 
If there is any preference suggested by the simulation data, it would be for the more negative dependence of retained metal fraction on stellar mass (e.g. fits R2 and R3), i.e. for more massive galaxies to expel a greater fraction of their metals than smaller ones, and correspondingly for there to be a turnover of the mass metallicity relation at high stellar masses, for example as exhibited in the HiZELS data in Fig.~\ref{fig:pred_metals}.

If the dependency of retained metal fraction on stellar mass were indeed this strong, one process which may reduce its effect is that of gravity. For massive galaxies metals may escape the disk but not the halo, leading to significant recycling of the metals. This could be parameterised either as a lower \lq effective\rq\ ejection fraction, or by re-introducing the metallicity $Z_{\rm a}$ of accreting gas in Eq. (\ref{eq:Mz_evol}) which was assumed to be zero in the analysis of Section \ref{sec:inf_metal}. One other intriguing possibility from this type of simulation is to study models of the radial distribution of metals in galaxy disks and compare to observational constraints e.g. from \cite{Kewley_2010}. We leave radial transport of metals (e.g. \citealp{Werk_2011}) to future work.

\section{Conclusions}\label{sec:conclusions}

In this paper we have performed a series of hydrodynamical simulations that extend the work of \CTB\ to trace the metal enrichment of the interstellar medium. We simulate patches of SN driven turbulence in a gravitationally bound disk, including cooling and metal enrichment via the SN ejecta, using the \flash\ code on a regular grid. Our simulations start from a column of gas which is initially in hydrostatic equilibrium. We assume that stars form at a rate set by the Kennicutt-Schmidt law, and that the associated supernovae inject mass, metals and energy into the ISM. The sub-parsec simulations trace the transition of the supernova ejecta from thermally driven to momentum driven, and chance overlap of hot SNe bubbles lead to regions in which the ISM is locally over-pressurised. These hot bubbles vent below and above the disk that in a time-averaged sense form a wind of hot, enriched gas, which is significantly mass loaded. We also experimented with including stellar winds, for example from the SN precursor or more generally from other massive stars that form together with the SN progenitor.

In order to make the simulations as scale-free as possible we used a simplified cooling function that allows us to consider the escape of the metal ejecta independently of the background metallicity of the ISM in individual runs, and allows easy comparison with the relations for metal ejection rates in \CTB. We confirmed that this does not have a strong impact on the results in Fig.~\ref{fig:metaldep_cool}, where we compared to simulations where metal-dependent cooling was enabled and alternate background metallicities were tried.

In Section \ref{sec:betaZResults} we investigated the metallicity of the ISM and the mixing produced by successive generations of SNe, where we found a metallicity bi-modality between an ejecta-rich hot phase, and a slowly enriching warm phase. 
This is primarily because SNe are the sites of both metal and energy injection. Given the molecular nature of diffusion, we expect metals and thermal energy to diffuse at the same rate implying a 
a metallicity - temperature relation (more enriched gas being hotter). 
This can be suppressed in the case of strong cooling, however this is not important for the hot phase ejected in the wind. 
For the cooler phases this correlation is correspondingly less strong, and would be further suppressed in our simulations had we included metal-dependent cooling.

In Section \ref{sec:met_outflow_deps} we parameterised the metal loss in the outflows as a function of gas surface density and gas fraction of the simulated columns. We found a weak dependence of metal mass loss on surface density, such that the metallicity of the outflow (the ratio of the metal mass loss to the total mass loss) has a negative dependence on surface density. In other words: {\em higher surface density disks have higher star formation rate, lower mass loading, with hotter, more enriched outflows}.

In Section \ref{sec:mass_metallicity} we constructed a simple model of the metallicity evolution of a galaxy, showing how the dependence of galaxy gas mass and metal ejection as a function of stellar mass can be used to infer the gas phase metallicities. 
The model includes inflow of gas, star formation, and gas and metal outflow which are assumed not to depend explicitly on time (redshift) yields a differential equation for the evolution of the ISM metallicity as function of stellar mass, which can be solved provided we have a model for how the gas fraction of a galaxy depends on its stellar mass.

By applying this model to observed fits to the $\Zg-M_\star$ gas metallicity - stellar mass relation we find that at low $M_\star$, the mass-metallicity relation is not set by an equilibrium but rather is determined by the dependence of the ratio of the gas mass to stellar mass, $\Mgas/M_\star$ on $M_\star$. In this case $\Zg$ is set by the cumulative yield of all the stars formed up to that time. A dependence $\Mgas/M_\star\propto M_\star^{0.5}$ (following \citealp{Peeples_2011}) yields $\Zg \propto M_\star^{0.5}$, consistent in normalisation and slope with the observed relation.
Higher $M_\star$ corresponds to lower $\Mgas$, with $\Zg$ now primarily determined by the instantaneous response of the gas reservoir to enrichment, yielding $\Zg$ nearly independent of $M_\star$. We explored what sets the turn-over location of the $\Zg-M_\star$ between these extremes.
Assuming galaxies have exponential disks, we can make use of our set of simulations at various gas surface densities to translate metal outflow rates as function of surface density, to outflow rates as function of {\em stellar mass}. This allows us to examine to what extent the {\em simulations} yield
metal retainment fractions that are consistent with the observed $\Zg-M_\star$ relation. The simulations yield retained metal fractions that {\em decrease} with $M_\star$, approximately $\propto M_\star^{-0.2}$ for models that assume that the gas fraction also decreases with increasing $M_\star$. Such a strong dependence does well in reproducing the observed $\Zg-M_\star$ relation.

Several avenues exist for future work, such as the modelling of the recycling of metals in halos to improve the constraints at high stellar masses. It would also be very interesting to apply this data set to study the radial evolution of metallicity in disk galaxies. The distribution of gas that exits these volumes is also of interest as it should form the basis of the CGM, and it may be possible to construct mock-absorption lines to compare to, for example, the cosmic origins spectrograph results.

\section*{Acknowledgements}

The authors would like to thank John Stott for sharing data prior to publication, and Mar\'ia-Emilia de Rossi for helpful comments. P.C. acknowledges support from the Leibniz Gemeinschaft through grant SAW-2012-AIP-5 129.
This work was supported by the Science and Technology Facilities Council (grant ST/F001166/1) and by the Interuniversity Attraction Poles Programme initiated by the Belgian Science Policy Office (grant IAP P7/08 CHARM).
This work used the DiRAC Data Centric system at Durham University, operated by the Institute for Computational Cosmology on behalf of the STFC DiRAC HPC Facility (www.dirac.ac.uk). This equipment was funded by BIS National E-infrastructure capital grant ST/K00042X/1, STFC capital grant ST/H008519/1, and STFC DiRAC Operations grant ST/K003267/1 and Durham University. DiRAC is part of the National E-Infrastructure. The \flash\ software used in this work was in part developed by the DOE-supported ASC/Alliance Center for Astrophysical Thermonuclear Flashes at the University of Chicago. The data generated in these simulations can be accessed through collaboration with the authors.

\bibliographystyle{mn2e}
\bibliography{metals} 

\begin{thebibliography}{89}
\expandafter\ifx\csname natexlab\endcsname\relax\def\natexlab#1{#1}\fi

\bibitem[{{Aguirre} {et~al.}(2005){Aguirre}, {Schaye}, {Hernquist}, {Kay},
  {Springel}, \& {Theuns}}]{Aguirre_05}
{Aguirre} A., {Schaye} J., {Hernquist} L., {Kay} S., {Springel} V., {Theuns}
  T., 2005, \apjl, 620, L13

\bibitem[{{Asplund} {et~al.}(2005){Asplund}, {Grevesse}, \&
  {Sauval}}]{Asplund_2005}
{Asplund} M., {Grevesse} N., {Sauval} A.~J., 2005, in Astronomical Society of
  the Pacific Conference Series, Vol. 336, Cosmic Abundances as Records of
  Stellar Evolution and Nucleosynthesis, {T.~G.~Barnes III \& F.~N.~Bash}, ed.,
  p.~25

\bibitem[{{Balogh} {et~al.}(2001){Balogh}, {Pearce}, {Bower}, \&
  {Kay}}]{Balogh_2001}
{Balogh} M.~L., {Pearce} F.~R., {Bower} R.~G., {Kay} S.~T., 2001, \mnras, 326,
  1228

\bibitem[{{Behroozi} {et~al.}(2010){Behroozi}, {Conroy}, \&
  {Wechsler}}]{Behroozi_10}
{Behroozi} P.~S., {Conroy} C., {Wechsler} R.~H., 2010, \apj, 717, 379

\bibitem[{{Bell} \& {de Jong}(2001)}]{Bell_2001}
{Bell} E.~F., {de Jong} R.~S., 2001, \apj, 550, 212

\bibitem[{{Benson} {et~al.}(2003){Benson}, {Bower}, {Frenk}, {Lacey}, {Baugh},
  \& {Cole}}]{Benson_2003}
{Benson} A.~J., {Bower} R.~G., {Frenk} C.~S., {Lacey} C.~G., {Baugh} C.~M.,
  {Cole} S., 2003, \apj, 599, 38

\bibitem[{{Bond}(1970)}]{Bond_1970}
{Bond} H.~E., 1970, \apjs, 22, 117

\bibitem[{{Brook} {et~al.}(2012){Brook}, {Stinson}, {Gibson}, {Wadsley}, \&
  {Quinn}}]{Brook_2012}
{Brook} C.~B., {Stinson} G., {Gibson} B.~K., {Wadsley} J., {Quinn} T., 2012,
  \mnras, 424, 1275

\bibitem[{{Burbidge} {et~al.}(1957){Burbidge}, {Burbidge}, {Fowler}, \&
  {Hoyle}}]{Burbidge_1957}
{Burbidge} E.~M., {Burbidge} G.~R., {Fowler} W.~A., {Hoyle} F., 1957, Reviews
  of Modern Physics, 29, 547

\bibitem[{{Castor} {et~al.}(1975){Castor}, {Abbott}, \& {Klein}}]{Castor_1975a}
{Castor} J.~I., {Abbott} D.~C., {Klein} R.~I., 1975, \apj, 195, 157

\bibitem[{{Catinella} {et~al.}(2010){Catinella}, {Schiminovich}, {Kauffmann},
  {Fabello}, {Wang}, {Hummels}, {Lemonias}, {Moran}, {Wu}, {Giovanelli},
  {Haynes}, {Heckman}, {Basu-Zych}, {Blanton}, {Brinchmann}, {Budav{\'a}ri},
  {Gon{\c c}alves}, {Johnson}, {Kennicutt}, {Madore}, {Martin}, {Rich},
  {Tacconi}, {Thilker}, {Wild}, \& {Wyder}}]{Catinella_2010}
{Catinella} B., {Schiminovich} D., {Kauffmann} G., {Fabello} S., {Wang} J.,
  {Hummels} C., {Lemonias} J., {Moran} S.~M., {Wu} R., {Giovanelli} R.,
  {Haynes} M.~P., {Heckman} T.~M., {Basu-Zych} A.~R., {Blanton} M.~R.,
  {Brinchmann} J., {Budav{\'a}ri} T., {Gon{\c c}alves} T., {Johnson} B.~D.,
  {Kennicutt} R.~C., {Madore} B.~F., {Martin} C.~D., {Rich} M.~R., {Tacconi}
  L.~J., {Thilker} D.~A., {Wild} V., {Wyder} T.~K., 2010, \mnras, 403, 683

\bibitem[{{Cen} {et~al.}(2005){Cen}, {Nagamine}, \& {Ostriker}}]{Cen_05}
{Cen} R., {Nagamine} K., {Ostriker} J.~P., 2005, \apj, 635, 86

\bibitem[{{Chabrier}(2003)}]{Chabrier_03}
{Chabrier} G., 2003, \pasp, 115, 763

\bibitem[{{Chevalier}(1974)}]{Chevalier_1974}
{Chevalier} R.~A., 1974, \apj, 188, 501

\bibitem[{{Cowie} {et~al.}(1995){Cowie}, {Songaila}, {Kim}, \&
  {Hu}}]{Cowie_1995}
{Cowie} L.~L., {Songaila} A., {Kim} T.-S., {Hu} E.~M., 1995, \aj, 109, 1522

\bibitem[{{Cox}(1972)}]{Cox_72}
{Cox} D.~P., 1972, \apj, 178, 159

\bibitem[{{Crain} {et~al.}(2010){Crain}, {McCarthy}, {Frenk}, {Theuns}, \&
  {Schaye}}]{Crain_2010}
{Crain} R.~A., {McCarthy} I.~G., {Frenk} C.~S., {Theuns} T., {Schaye} J., 2010,
  \mnras, 407, 1403

\bibitem[{{Creasey} {et~al.}(2013){Creasey}, {Theuns}, \& {Bower}}]{Creasey_13}
{Creasey} P., {Theuns} T., {Bower} R.~G., 2013, \mnras, 429, 1922

\bibitem[{{Dav{\'e}} {et~al.}(2011){Dav{\'e}}, {Finlator}, \&
  {Oppenheimer}}]{Dave_2011}
{Dav{\'e}} R., {Finlator} K., {Oppenheimer} B.~D., 2011, \mnras, 416, 1354

\bibitem[{{Dav{\'e}} {et~al.}(2012){Dav{\'e}}, {Finlator}, \&
  {Oppenheimer}}]{Dave_2012}
---, 2012, \mnras, 421, 98

\bibitem[{{De Rossi} {et~al.}(2013){De Rossi}, {Avila-Reese}, {Tissera},
  {Gonz{\'a}lez-Samaniego}, \& {Pedrosa}}]{deRossi_2013}
{De Rossi} M.~E., {Avila-Reese} V., {Tissera} P.~B., {Gonz{\'a}lez-Samaniego}
  A., {Pedrosa} S.~E., 2013, \mnras, 435, 2736

\bibitem[{{Denicol{\'o}} {et~al.}(2002){Denicol{\'o}}, {Terlevich}, \&
  {Terlevich}}]{Denicolo_2002}
{Denicol{\'o}} G., {Terlevich} R., {Terlevich} E., 2002, \mnras, 330, 69

\bibitem[{{Edmunds}(1990)}]{Edmunds_1990}
{Edmunds} M.~G., 1990, \mnras, 246, 678

\bibitem[{{Elmegreen} \& {Scalo}(2004)}]{Elmegreen_2004}
{Elmegreen} B.~G., {Scalo} J., 2004, \araa, 42, 211

\bibitem[{{Ferrara} {et~al.}(2000){Ferrara}, {Pettini}, \&
  {Shchekinov}}]{Ferrara_2000}
{Ferrara} A., {Pettini} M., {Shchekinov} Y., 2000, \mnras, 319, 539

\bibitem[{{Ferrara} {et~al.}(2005){Ferrara}, {Scannapieco}, \&
  {Bergeron}}]{Ferrara_2005}
{Ferrara} A., {Scannapieco} E., {Bergeron} J., 2005, \apjl, 634, L37

\bibitem[{{Finlator} \& {Dav{\'e}}(2008)}]{Finlator_2008}
{Finlator} K., {Dav{\'e}} R., 2008, \mnras, 385, 2181

\bibitem[{{Fryxell} {et~al.}(2000){Fryxell}, {Olson}, {Ricker}, {Timmes},
  {Zingale}, {Lamb}, {MacNeice}, {Rosner}, {Truran}, \& {Tufo}}]{Fryxell00}
{Fryxell} B., {Olson} K., {Ricker} P., {Timmes} F.~X., {Zingale} M., {Lamb}
  D.~Q., {MacNeice} P., {Rosner} R., {Truran} J.~W., {Tufo} H., 2000, \apjs,
  131, 273

\bibitem[{{Fukugita} {et~al.}(1998){Fukugita}, {Hogan}, \&
  {Peebles}}]{Fukugita_98}
{Fukugita} M., {Hogan} C.~J., {Peebles} P.~J.~E., 1998, \apj, 503, 518

\bibitem[{{Gent} {et~al.}(2013){Gent}, {Shukurov}, {Fletcher}, {Sarson}, \&
  {Mantere}}]{Gent_2012}
{Gent} F.~A., {Shukurov} A., {Fletcher} A., {Sarson} G.~R., {Mantere} M.~J.,
  2013, \mnras, 432, 1396

\bibitem[{{Guo} {et~al.}(2010){Guo}, {White}, {Li}, \&
  {Boylan-Kolchin}}]{Guo_10}
{Guo} Q., {White} S., {Li} C., {Boylan-Kolchin} M., 2010, \mnras, 404, 1111

\bibitem[{{Heckman} {et~al.}(1990){Heckman}, {Armus}, \&
  {Miley}}]{Heckman_1990}
{Heckman} T.~M., {Armus} L., {Miley} G.~K., 1990, \apjs, 74, 833

\bibitem[{{Heckman} {et~al.}(2000){Heckman}, {Lehnert}, {Strickland}, \&
  {Armus}}]{Heckman_2000}
{Heckman} T.~M., {Lehnert} M.~D., {Strickland} D.~K., {Armus} L., 2000, \apjs,
  129, 493

\bibitem[{{Hill} {et~al.}(2012){Hill}, {Joung}, {Mac Low}, {Benjamin},
  {Haffner}, {Klingenberg}, \& {Waagan}}]{Hill_2012}
{Hill} A.~S., {Joung} M.~R., {Mac Low} M.-M., {Benjamin} R.~A., {Haffner}
  L.~M., {Klingenberg} C., {Waagan} K., 2012, \apj, 750, 104

\bibitem[{{Holmberg} \& {Flynn}(2004)}]{Holmberg_2004}
{Holmberg} J., {Flynn} C., 2004, \mnras, 352, 440

\bibitem[{{Hopkins} {et~al.}(2011){Hopkins}, {Quataert}, \&
  {Murray}}]{Hopkins_2011a}
{Hopkins} P.~F., {Quataert} E., {Murray} N., 2011, \mnras, 417, 950

\bibitem[{{Hopkins} {et~al.}(2012){Hopkins}, {Quataert}, \&
  {Murray}}]{Hopkins_2012}
---, 2012, \mnras, 421, 3488

\bibitem[{{J{\o}rgensen}(2000)}]{Jorgensen_2000}
{J{\o}rgensen} B.~R., 2000, \aap, 363, 947

\bibitem[{{Joung} \& {Mac Low}(2006)}]{Joung_MacLow_2006}
{Joung} M.~K.~R., {Mac Low} M.-M., 2006, \apj, 653, 1266

\bibitem[{{Kennicutt}(1998)}]{Kennicutt_98}
{Kennicutt} Jr. R.~C., 1998, \araa, 36, 189

\bibitem[{{Kewley} \& {Ellison}(2008)}]{Kewley_2008}
{Kewley} L.~J., {Ellison} S.~L., 2008, \apj, 681, 1183

\bibitem[{{Kewley} {et~al.}(2010){Kewley}, {Rupke}, {Zahid}, {Geller}, \&
  {Barton}}]{Kewley_2010}
{Kewley} L.~J., {Rupke} D., {Zahid} H.~J., {Geller} M.~J., {Barton} E.~J.,
  2010, \apjl, 721, L48

\bibitem[{{Larson}(1974)}]{Larson_1974}
{Larson} R.~B., 1974, \mnras, 169, 229

\bibitem[{{Leroy} {et~al.}(2008){Leroy}, {Walter}, {Brinks}, {Bigiel}, {de
  Blok}, {Madore}, \& {Thornley}}]{Leroy_2008}
{Leroy} A.~K., {Walter} F., {Brinks} E., {Bigiel} F., {de Blok} W.~J.~G.,
  {Madore} B., {Thornley} M.~D., 2008, \aj, 136, 2782

\bibitem[{{Martin}(2005)}]{Martin_2005}
{Martin} C.~L., 2005, \apj, 621, 227

\bibitem[{{Martin} {et~al.}(2013){Martin}, {Shapley}, {Coil}, {Kornei},
  {Murray}, \& {Pancoast}}]{Martin_13}
{Martin} C.~L., {Shapley} A.~E., {Coil} A.~L., {Kornei} K.~A., {Murray} N.,
  {Pancoast} A., 2013, \apj, 770, 41

\bibitem[{{McGaugh}(2005)}]{McGaugh_2005}
{McGaugh} S.~S., 2005, \apj, 632, 859

\bibitem[{{McKee} \& {Ostriker}(1977)}]{McKee_Ostriker_1977}
{McKee} C.~F., {Ostriker} J.~P., 1977, \apj, 218, 148

\bibitem[{{Meiring} {et~al.}(2013){Meiring}, {Tripp}, {Werk}, {Howk},
  {Jenkins}, {Prochaska}, {Lehner}, \& {Sembach}}]{Werk_2012}
{Meiring} J.~D., {Tripp} T.~M., {Werk} J.~K., {Howk} J.~C., {Jenkins} E.~B.,
  {Prochaska} J.~X., {Lehner} N., {Sembach} K.~R., 2013, \apj, 767, 49

\bibitem[{{Mori} {et~al.}(2002){Mori}, {Ferrara}, \& {Madau}}]{Mori_02}
{Mori} M., {Ferrara} A., {Madau} P., 2002, \apj, 571, 40

\bibitem[{{Murray} {et~al.}(2005){Murray}, {Quataert}, \&
  {Thompson}}]{Murray_2005}
{Murray} N., {Quataert} E., {Thompson} T.~A., 2005, \apj, 618, 569

\bibitem[{{Nordstr{\"o}m} {et~al.}(2004){Nordstr{\"o}m}, {Mayor}, {Andersen},
  {Holmberg}, {Pont}, {J{\o}rgensen}, {Olsen}, {Udry}, \&
  {Mowlavi}}]{Nordstrom_2004}
{Nordstr{\"o}m} B., {Mayor} M., {Andersen} J., {Holmberg} J., {Pont} F.,
  {J{\o}rgensen} B.~R., {Olsen} E.~H., {Udry} S., {Mowlavi} N., 2004, \aap,
  418, 989

\bibitem[{{Pagel} \& {Patchett}(1975)}]{Pagel_1975}
{Pagel} B.~E.~J., {Patchett} B.~E., 1975, \mnras, 172, 13

\bibitem[{{Papastergis} {et~al.}(2012){Papastergis}, {Cattaneo}, {Huang},
  {Giovanelli}, \& {Haynes}}]{Pap_2012}
{Papastergis} E., {Cattaneo} A., {Huang} S., {Giovanelli} R., {Haynes} M.~P.,
  2012, \apj, 759, 138

\bibitem[{{Peeples} \& {Shankar}(2011)}]{Peeples_2011}
{Peeples} M.~S., {Shankar} F., 2011, \mnras, 417, 2962

\bibitem[{{Peeples} {et~al.}(2014){Peeples}, {Werk}, {Tumlinson},
  {Oppenheimer}, {Prochaska}, {Katz}, \& {Weinberg}}]{Peeples_2014}
{Peeples} M.~S., {Werk} J.~K., {Tumlinson} J., {Oppenheimer} B.~D., {Prochaska}
  J.~X., {Katz} N., {Weinberg} D.~H., 2014, \apj, 786, 54

\bibitem[{{Pettini} {et~al.}(2001){Pettini}, {Shapley}, {Steidel}, {Cuby},
  {Dickinson}, {Moorwood}, {Adelberger}, \& {Giavalisco}}]{Pettini_2001}
{Pettini} M., {Shapley} A.~E., {Steidel} C.~C., {Cuby} J.-G., {Dickinson} M.,
  {Moorwood} A.~F.~M., {Adelberger} K.~L., {Giavalisco} M., 2001, \apj, 554,
  981

\bibitem[{{Pilkington} {et~al.}(2012){Pilkington}, {Gibson}, {Brook}, {Calura},
  {Stinson}, {Thacker}, {Michel-Dansac}, {Bailin}, {Couchman}, {Wadsley},
  {Quinn}, \& {Maccio}}]{Pilkington_2012}
{Pilkington} K., {Gibson} B.~K., {Brook} C.~B., {Calura} F., {Stinson} G.~S.,
  {Thacker} R.~J., {Michel-Dansac} L., {Bailin} J., {Couchman} H.~M.~P.,
  {Wadsley} J., {Quinn} T.~R., {Maccio} A., 2012, \mnras, 3484

\bibitem[{{Rees} \& {Ostriker}(1977)}]{Rees_1977}
{Rees} M.~J., {Ostriker} J.~P., 1977, \mnras, 179, 541

\bibitem[{{Rosdahl} {et~al.}(2013){Rosdahl}, {Blaizot}, {Aubert}, {Stranex}, \&
  {Teyssier}}]{Rosdahl_2013}
{Rosdahl} J., {Blaizot} J., {Aubert} D., {Stranex} T., {Teyssier} R., 2013,
  \mnras, 436, 2188

\bibitem[{{Schaye} {et~al.}(2003){Schaye}, {Aguirre}, {Kim}, {Theuns}, {Rauch},
  \& {Sargent}}]{Schaye_2003}
{Schaye} J., {Aguirre} A., {Kim} T.-S., {Theuns} T., {Rauch} M., {Sargent}
  W.~L.~W., 2003, \apj, 596, 768

\bibitem[{{Schaye} {et~al.}(2014){Schaye}, {Crain}, {Bower}, {Furlong},
  {Schaller}, {Theuns}, {Dalla Vecchia}, {Frenk}, {McCarthy}, {Helly},
  {Jenkins}, {Rosas-Guevara}, {White}, {Baes}, {Booth}, {Camps}, {Navarro},
  {Qu}, {Rahmati}, {Sawala}, {Thomas}, \& {Trayford}}]{Schaye_2014}
{Schaye} J., {Crain} R.~A., {Bower} R.~G., {Furlong} M., {Schaller} M.,
  {Theuns} T., {Dalla Vecchia} C., {Frenk} C.~S., {McCarthy} I.~G., {Helly}
  J.~C., {Jenkins} A., {Rosas-Guevara} Y.~M., {White} S.~D.~M., {Baes} M.,
  {Booth} C.~M., {Camps} P., {Navarro} J.~F., {Qu} Y., {Rahmati} A., {Sawala}
  T., {Thomas} P.~A., {Trayford} J., 2014, ArXiv e-prints, arXiv:1407.7040

\bibitem[{{Schaye} {et~al.}(2010){Schaye}, {Dalla Vecchia}, {Booth}, {Wiersma},
  {Theuns}, {Haas}, {Bertone}, {Duffy}, {McCarthy}, \& {van de
  Voort}}]{Schaye_10}
{Schaye} J., {Dalla Vecchia} C., {Booth} C.~M., {Wiersma} R.~P.~C., {Theuns}
  T., {Haas} M.~R., {Bertone} S., {Duffy} A.~R., {McCarthy} I.~G., {van de
  Voort} F., 2010, \mnras, 402, 1536

\bibitem[{{Schmidt}(1963)}]{Schmidt_1963}
{Schmidt} M., 1963, \apj, 137, 758

\bibitem[{{Sharma} {et~al.}(2011){Sharma}, {Nath}, \&
  {Shchekinov}}]{Sharma_2011}
{Sharma} M., {Nath} B.~B., {Shchekinov} Y., 2011, \apjl, 736, L27

\bibitem[{{Shen} {et~al.}(2003){Shen}, {Mo}, {White}, {Blanton}, {Kauffmann},
  {Voges}, {Brinkmann}, \& {Csabai}}]{Shen_2003}
{Shen} S., {Mo} H.~J., {White} S.~D.~M., {Blanton} M.~R., {Kauffmann} G.,
  {Voges} W., {Brinkmann} J., {Csabai} I., 2003, \mnras, 343, 978

\bibitem[{{Slyz} {et~al.}(2005){Slyz}, {Devriendt}, {Bryan}, \&
  {Silk}}]{Slyz_2005}
{Slyz} A.~D., {Devriendt} J.~E.~G., {Bryan} G., {Silk} J., 2005, \mnras, 356,
  737

\bibitem[{{Springel} \& {Hernquist}(2003)}]{Springel_03}
{Springel} V., {Hernquist} L., 2003, \mnras, 339, 289

\bibitem[{{Stott} {et~al.}(2013){Stott}, {Sobral}, {Bower}, {Smail}, {Best},
  {Matsuda}, {Hayashi}, {Geach}, \& {Kodama}}]{Stott_2013}
{Stott} J.~P., {Sobral} D., {Bower} R., {Smail} I., {Best} P.~N., {Matsuda} Y.,
  {Hayashi} M., {Geach} J.~E., {Kodama} T., 2013, \mnras, 436, 1130

\bibitem[{{Sutherland} \& {Dopita}(1993)}]{Sutherland_1993}
{Sutherland} R.~S., {Dopita} M.~A., 1993, \apjs, 88, 253

\bibitem[{{Theuns} {et~al.}(2002){Theuns}, {Viel}, {Kay}, {Schaye}, {Carswell},
  \& {Tzanavaris}}]{Theuns_02}
{Theuns} T., {Viel} M., {Kay} S., {Schaye} J., {Carswell} R.~F., {Tzanavaris}
  P., 2002, \apjl, 578, L5

\bibitem[{{Tinsley}(1980)}]{Tinsley_1980}
{Tinsley} B.~M., 1980, \fcp, 5, 287

\bibitem[{{Tremonti} {et~al.}(2004){Tremonti}, {Heckman}, {Kauffmann},
  {Brinchmann}, {Charlot}, {White}, {Seibert}, {Peng}, {Schlegel}, {Uomoto},
  {Fukugita}, \& {Brinkmann}}]{Tremonti_2004}
{Tremonti} C.~A., {Heckman} T.~M., {Kauffmann} G., {Brinchmann} J., {Charlot}
  S., {White} S.~D.~M., {Seibert} M., {Peng} E.~W., {Schlegel} D.~J., {Uomoto}
  A., {Fukugita} M., {Brinkmann} J., 2004, \apj, 613, 898

\bibitem[{{Tumlinson} {et~al.}(2013){Tumlinson}, {Thom}, {Werk}, {Prochaska},
  {Tripp}, {Katz}, {Dav{\'e}}, {Oppenheimer}, {Meiring}, {Ford}, {O'Meara},
  {Peeples}, {Sembach}, \& {Weinberg}}]{Tumlinson_2013}
{Tumlinson} J., {Thom} C., {Werk} J.~K., {Prochaska} J.~X., {Tripp} T.~M.,
  {Katz} N., {Dav{\'e}} R., {Oppenheimer} B.~D., {Meiring} J.~D., {Ford} A.~B.,
  {O'Meara} J.~M., {Peeples} M.~S., {Sembach} K.~R., {Weinberg} D.~H., 2013,
  \apj, 777, 59

\bibitem[{{Tumlinson} {et~al.}(2011){Tumlinson}, {Thom}, {Werk}, {Prochaska},
  {Tripp}, {Weinberg}, {Peeples}, {O'Meara}, {Oppenheimer}, {Meiring}, {Katz},
  {Dav{\'e}}, {Ford}, \& {Sembach}}]{Tumlinson_2011}
{Tumlinson} J., {Thom} C., {Werk} J.~K., {Prochaska} J.~X., {Tripp} T.~M.,
  {Weinberg} D.~H., {Peeples} M.~S., {O'Meara} J.~M., {Oppenheimer} B.~D.,
  {Meiring} J.~D., {Katz} N.~S., {Dav{\'e}} R., {Ford} A.~B., {Sembach} K.~R.,
  2011, Science, 334, 948

\bibitem[{{van de Voort} {et~al.}(2012){van de Voort}, {Schaye}, {Altay}, \&
  {Theuns}}]{deVoort_2012}
{van de Voort} F., {Schaye} J., {Altay} G., {Theuns} T., 2012, \mnras, 421,
  2809

\bibitem[{{van den Bergh}(1962)}]{van_den_Bergh_1962}
{van den Bergh} S., 1962, \aj, 67, 486

\bibitem[{{Walter} {et~al.}(2008){Walter}, {Brinks}, {de Blok}, {Bigiel},
  {Kennicutt}, {Thornley}, \& {Leroy}}]{Walter_2008}
{Walter} F., {Brinks} E., {de Blok} W.~J.~G., {Bigiel} F., {Kennicutt} Jr.
  R.~C., {Thornley} M.~D., {Leroy} A., 2008, \aj, 136, 2563

\bibitem[{{Werk} {et~al.}(2014){Werk}, {Prochaska}, {Tumlinson}, {Peeples},
  {Tripp}, {Fox}, {Lehner}, {Thom}, {O'Meara}, {Ford}, {Bordoloi}, {Katz},
  {Tejos}, {Oppenheimer}, {Dav{\'e}}, \& {Weinberg}}]{Werk_2014}
{Werk} J.~K., {Prochaska} J.~X., {Tumlinson} J., {Peeples} M.~S., {Tripp}
  T.~M., {Fox} A.~J., {Lehner} N., {Thom} C., {O'Meara} J.~M., {Ford} A.~B.,
  {Bordoloi} R., {Katz} N., {Tejos} N., {Oppenheimer} B.~D., {Dav{\'e}} R.,
  {Weinberg} D.~H., 2014, ArXiv e-prints

\bibitem[{{Werk} {et~al.}(2011){Werk}, {Putman}, {Meurer}, \&
  {Santiago-Figueroa}}]{Werk_2011}
{Werk} J.~K., {Putman} M.~E., {Meurer} G.~R., {Santiago-Figueroa} N., 2011,
  \apj, 735, 71

\bibitem[{{West} {et~al.}(2009){West}, {Garcia-Appadoo}, {Dalcanton}, {Disney},
  {Rockosi}, \& {Ivezi{\'c}}}]{West_2009}
{West} A.~A., {Garcia-Appadoo} D.~A., {Dalcanton} J.~J., {Disney} M.~J.,
  {Rockosi} C.~M., {Ivezi{\'c}} {\v Z}., 2009, \aj, 138, 796

\bibitem[{{White} \& {Rees}(1978)}]{WhiteRees_1978}
{White} S.~D.~M., {Rees} M.~J., 1978, \mnras, 183, 341

\bibitem[{{Wiersma} {et~al.}(2009{\natexlab{a}}){Wiersma}, {Schaye}, \&
  {Smith}}]{Wiersma_Schaye_and_Smith_09}
{Wiersma} R.~P.~C., {Schaye} J., {Smith} B.~D., 2009{\natexlab{a}}, \mnras, 20

\bibitem[{{Wiersma} {et~al.}(2009{\natexlab{b}}){Wiersma}, {Schaye}, {Theuns},
  {Dalla Vecchia}, \& {Tornatore}}]{Wiersma_09}
{Wiersma} R.~P.~C., {Schaye} J., {Theuns} T., {Dalla Vecchia} C., {Tornatore}
  L., 2009{\natexlab{b}}, \mnras, 399, 574

\bibitem[{{Wise} {et~al.}(2012){Wise}, {Abel}, {Turk}, {Norman}, \&
  {Smith}}]{Wise_12}
{Wise} J.~H., {Abel} T., {Turk} M.~J., {Norman} M.~L., {Smith} B.~D., 2012,
  \mnras, 427, 311

\bibitem[{{Woosley} {et~al.}(1973){Woosley}, {Arnett}, \&
  {Clayton}}]{Woosley_73}
{Woosley} S.~E., {Arnett} W.~D., {Clayton} D.~D., 1973, \apjs, 26, 231

\bibitem[{{Woosley} \& {Weaver}(1995)}]{Woosley_1995}
{Woosley} S.~E., {Weaver} T.~A., 1995, \apjs, 101, 181

\bibitem[{{Worthey}(1994)}]{Worthey_1994}
{Worthey} G., 1994, \apjs, 95, 107

\bibitem[{{Zahid} {et~al.}(2014){Zahid}, {Dima}, {Kudritzki}, {Kewley},
  {Geller}, {Hwang}, {Silverman}, \& {Kashino}}]{Zahid_2014}
{Zahid} J., {Dima} G., {Kudritzki} R., {Kewley} L., {Geller} M., {Hwang} H.~S.,
  {Silverman} J., {Kashino} D., 2014, ArXiv e-prints

\end{thebibliography}
\bsp

\begin{appendix}

\section{Comparison with closed box models}\label{sec:closedbox}
\begin{figure}
\centering
\includegraphics[width=\figwidth]{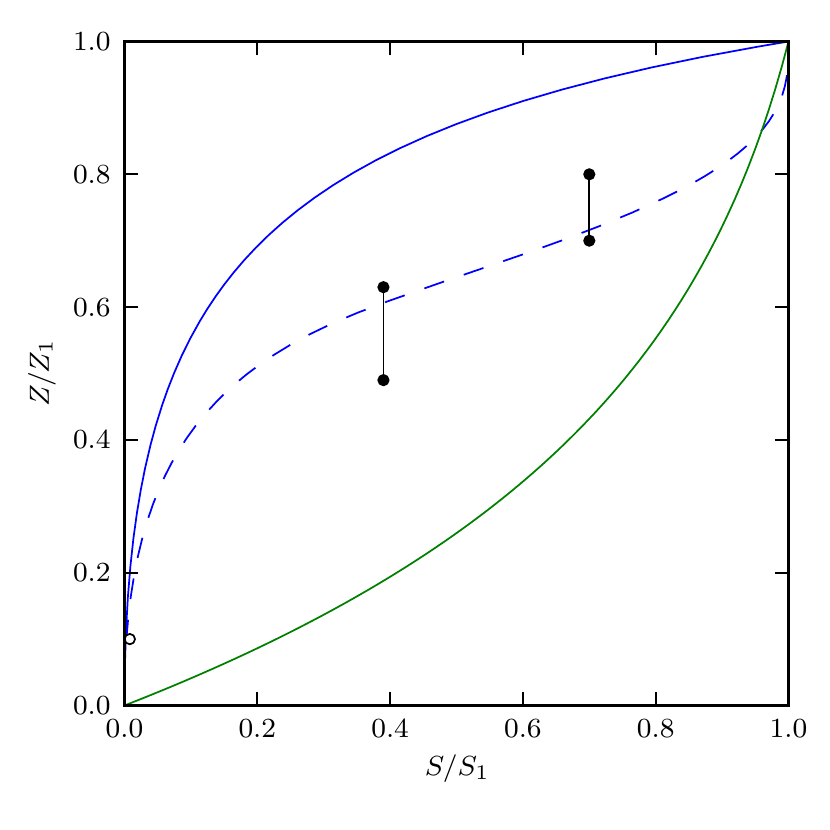}
\caption[Cumulative metallicity distribution faint stars]{Cumulative normalised metallicity distribution of faint stars, the horizontal axis indicating the proportion of stars below the metallicity indicated on the vertical axis. \emph{Green line} is the closed box model with $\mu_1=0.2$, \emph{blue line} is the simple metallicity evolution described in the text, \emph{blue dashed line} is the same model but assuming the faint stars form with a scatter in metallicity of $0.06$ dex. \emph{Solid black circles} are the data from \cite{Schmidt_1963} and the \emph{empty circle} is the data point from \cite{Bond_1970}.}
\label{fig:gdwarfs}
\end{figure}

One of the simplest ways to test the metal enrichment scenario described in Sect.~\ref{sec:mass_metallicity} is to estimate the number of old faint stars of a given metallicity. Lifetimes of sufficiently low-mass stars exceed the age of the universe and thus they become a tracer of the evolution of star formation. The relative absence of low metallicity faint stars (\citealp{van_den_Bergh_1962} and \citealp{Schmidt_1963}) compared to that predicted by the \lq closed box\rq\ model (described next) has become colloquially known as \lq the G-Dwarf problem\rq. 

The closed box model of star formation assumes that there are two types of stars, those with low and high masses. A given initial mass of gas gets converted to stars, without mass loss from or mass accretion into the box. The high mass stars explode immediately and return enriched gas to the ISM, whilst the low mass stars lock away their progenitor ISM metallicity indefinitely. The model further assumes the ISM to be completely homogeneous in metallicity. Assuming an ISM gas reservoir of mass $\Mgas(t)$ that is converted into stars, the metallicity of this gas reservoir will be 
\begin{equation}\label{eq:closed_box}
\Zg = -y \ln \mu \, ,
\end{equation}
where $\mu = \Mgas / (M_\star + \Mgas)$ is the gas to total mass fraction which gradually becomes more polluted (metal rich) as the remaining gas reservoir depletes, i.e. the gas metallicity is a monotonically decreasing function of $\mu$.

The distribution function of stars of different metallicities in this model is also a monotonically decreasing function of metallicity \citep{Schmidt_1963,Pagel_1975,Edmunds_1990} and the cumulative distribution (traditionally but somewhat awkwardly defined as the inverse cumulative distribution function of metallicity fraction as a function of stellar fraction, due to \citealp{Schmidt_1963}) will be a convex function with a large tail of low metallicity faint stars, i.e.
\begin{equation}
\frac{S(< Z) }{S(<Z_1)} = \frac{1-\mu_1^{Z/Z_1} }{1-\mu_1} \, ,
\end{equation}
where $S(<Z)$ denotes the number of stars of metallicity $<Z$, and $Z_1, \mu_1$ are the maximum (minimum) metallicity (gas fraction) in the closed box model, related by Eq. (\ref{eq:closed_box}). The observations (e.g. \citealp{Bond_1970}), however, do not find these low metallicity stars and suggest that the distribution will be largely concave\footnote{If the distribution function looks even close to Gaussian then the inverse cumulative distribution will always be concave at low fractions and convex at high ones, so they are unlikely to be described as entirely concave or convex.}.

In order to produce a peak in the distribution function it is necessary that the accelerating rise of metallicity of the ISM be stalled at some stage. This is usually understood to require inflow (e.g. \citealp{Edmunds_1990}), as a model with outflow, although removing metals, cannot reduce the \emph{mean} metallicity of a homogeneous ISM, which would need some low metallicity inflow to dilute. As we have seen, however, the outflows in our simulations are of higher metallicity than the average of the ISM. They preferentially remove high metallicity gas and so deplete the average metallicity of the ISM.

With this in mind we applied the corresponding approximations to calculate the metallicity of stars using the formalism of Section~\ref{sec:mass_metallicity} . In Fig.~\ref{fig:gdwarfs} we show the metallicity distributions of faint stars as predicted by the closed box model, where the fraction of gas to total mass ($\mu_1$) is $0.2$, and observational points found by \cite{Schmidt_1963} and \cite{Bond_1970}. We also show the result of integrating Eq.~(\ref{eq:Zg_evol_mstar}) to find the metallicity distribution of stars for a $M_\star = 3\times 10^{10} \; \rm \Msun$ galaxy with gas evolution given by Eq. (\ref{eq:mgass_mstar}) with $A_{\rm g}=1$ and $\alpha_{\rm g}=1/2$. More recent data exists which we will discuss shortly, here we are simply contrasting the models.

The closed box model predicts that nearly 40 per cent of stars have metallicity below 20 per cent of the maximum, that is -- a large number of low metallicity \lq G\rq-dwarfs -- that is not seen in the data, the so-called G-dwarf problem. To resolve the discrepancy, \cite{Schmidt_1963}  introduced a time dependent initial mass function to produce more massive stars at early times in the galaxy's evolution. These stars enrich the ISM without leaving low-metallicity remnants. The model from  Eq.~(\ref{eq:Zg_evol_mstar}) does a very good job of keeping the number of low metallicity stars small, but seems to have a normalisation problem against the Schmidt data. This, however, is the result of taking the fraction of the maximum metallicity. Eq.~(\ref{eq:Zg_evol_mstar}) has a sharp cut-off at a maximum metallicity of $\Zg(M_\star )$ which carries through to the cumulative distribution. A more realistic model would have the faint stars forming with some scatter about the mean metallicity of the ISM, producing a tail at higher metallicity which would lower the normalisation in Fig.~\ref{fig:gdwarfs} to closer agreement with the data, and we have illustrated the effect of this by showing the same model with $\pm 0.06$ dex of scatter.

\begin{figure}
\centering
\includegraphics[width=\figwidth]{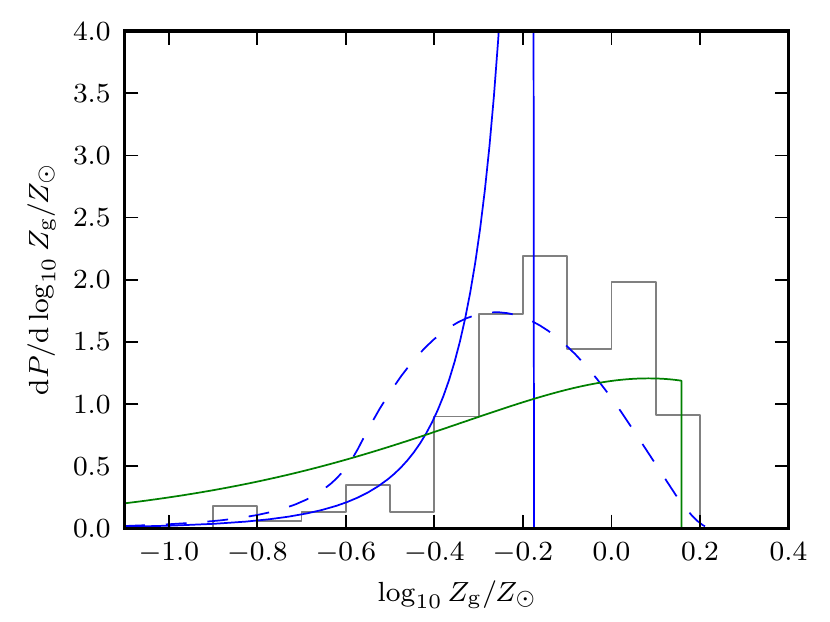}
\caption[Probability density function of faint stars.]{Probability density function of faint stars. \emph{Grey histogram} is the data from \cite{Jorgensen_2000}. \emph{Solid blue line} is the metallicity evolution model described in the text, \emph{dotted blue line} is the same model but with $\pm 0.18$ dex in scatter, \emph{green line} is the equivalent closed box model, which suffers from the G-dwarf problem.}
\label{fig:gdwarfs_pdf}
\end{figure}

To avoid the problems of normalisation due to a tail in high metallicity stars, a more robust method is to calculate the PDF of stellar metallicities. In Fig.~\ref{fig:gdwarfs_pdf} we also compare to the data of \cite{Jorgensen_2000} for stars in the range $0.7 < {\rm M/\Msun} < 1.0$. More extensive Hipparcos data was analysed in \cite{Nordstrom_2004}, with very similar distribution. After the addition of $0.18$ dex scatter in metallicity the model shows a good agreement in profile to the data, but with an offset in metallicity where the data is approximately $0.2$ dex higher.  One way to achieve this would be to slightly increase the yield or the retained metal fractions by this factor. For reference we also show the closed box model for the same galaxy, which reaches higher metallicity (since it does not lose metals) and has the unobserved low metallicity population.

\label{lastpage}
\end{appendix}

\end{document}